\documentclass[a4paper]{article}
\usepackage{amsmath}
\usepackage{amssymb}
\usepackage{amsfonts}
\usepackage{amsthm}
\usepackage{amsopn}
\usepackage{graphicx}
\usepackage{tikz}
\usepackage{url}
\usepackage{float}
\usepackage{hyphenat}
\usepackage[normalem]{ulem}

\def\be{\begin{eqnarray}}
\def\ee{\end{eqnarray}}
\def\nn{\nonumber}
\def\lb{\left (}
\def\rb{\right )}

\def\l[{\phantom.[}

\theoremstyle{plain}                    %stile corsivo
\newtheorem{teo}{Theorem}[section]      %definizione ambiente teorema
\newtheorem{conj}{Conjecture}[section]      %definizione ambiente teorema
\newtheorem{prop}[teo]{Proposition}    %definizione ambiente proposizione
\theoremstyle{definition}               %stile roman
\theoremstyle{remark}                   %stile per osservazioni
\newtheorem{rmk}{Remark}           %definizione ambiente osservazione

\hoffset= - 1.0in         % switch off for draft style
\begin{document}

\title{{\bf {
%The rich world of virtual knots
On Ambiguity in Knot Polynomials for Virtual Knots
}\vspace{.2cm}}
\author{{\bf A.Morozov$^{a,b,c}$},\ {\bf And.Morozov$^{a,b,c,d}$} \ and \ {\bf A.Popolitov$^{a,b,e}$}}
\date{ }
}

\maketitle

\vspace{-4.5cm}

\begin{center}
\hfill IITP/TH-18/15
\end{center}

\vspace{3.3cm}

\begin{center}

$^a$ {\small {\it ITEP, Moscow 117218, Russia}}\\
$^b$ {\small {\it Institute for Information Transmission Problems, Moscow 127994, Russia}}\\
$^c$ {\small {\it National Research Nuclear University MEPhI, Moscow 115409, Russia }}\\
$^d$ {\small {\it Laboratory of Quantum Topology, Chelyabinsk State University, Chelyabinsk 454001, Russia}} \\
$^e$ {\small {\it Korteweg-de Vries Institute for Mathematics, University of
Amsterdam, \\P.O. Box 94248, 1090 GE Amsterdam, The Netherlands}}
\end{center}

%\vspace{3.7cm}

\vspace{1cm}

\centerline{ABSTRACT}

\bigskip

{\footnotesize
  We claim that HOMFLY polynomials for virtual knots, defined with
  the help of the matrix-model recursion relations,
  contain more parameters, than just the usual $q$ and $A = q^N$.
  These parameters preserve topological invariance and do not show up in the case of ordinary
  (non-virtual) knots and links. They are most conveniently observed in the hypercube formalism:
  then they substitute $q$-dimensions of certain fat graphs,
  which are not constrained by recursion and can be chosen arbitrarily.
  The number of these new topological invariants seems to grow fast with
  the number of non-virtual crossings: 0, 1, 1, 5, 15, 91, 784, 9160, ...
  This number can be decreased by imposing the factorization requirement for composites,
  in addition to topological invariance -- still freedom remains.
  None of these new parameters, however, appear in HOMFLY for Kishino unknot,
  which thus remains unseparated from the ordinary unknots even by this enriched set of knot invariants.
}

\bigskip

\bigskip

\section{Introduction}

The most efficient methods to calculate knot/link polynomials \cite{knotpols,superpols}
are actually dealing with link diagrams -- oriented graphs of valence $(2,2)$
with two types (colors) of vertices.
In the  Chern-Simons theory \cite{CS} this description appears in
the temporal gauge $A_0=0$ \cite{MorSmi} (while a choice of holomorphic gauge
$A_1+iA_2=0$ leads to the formalism of Kontsevich integrals \cite{KoI}).
Topological invariance is then provided by the Reidemeister invariance.
For ordinary knots and links graphs are planar, non-planar graphs are
interpreted as associated with {\it virtual} knots/links,
provided the set of Reidemeister moves is appropriately enlarged \cite{Kvirt}.
Planar graphs are intimately related to braids, what allows to apply the
powerful theory of Hecke algebras -- which is in the base of the modern
(universal) versions \cite{RTmod,RTnew} of the Reshetikhin-Turaev (RT) formalism \cite{RT}.
This is not the case for non-planar graphs, and these methods are not
directly applicable
%-- though significant pieces of RT formalism,
%like skein relations in the fundamental representation, survive and
(though available {\it answers} imply that just a minor --
but still unknown -- modification can be needed).
Therefore virtual knot/link polynomials need to be studied by different methods.
Historically first was the artful skein relation method \cite{Kvirt}
(which is normally a small part of the RT formalism, restricted to the fundamental
representation \cite{RTmod}).
A more systematic alternative approach is the hypercube method of \cite{BN,DM3},
which was successfully applied to virtual knots in \cite{virtHOMFLY} and \cite{MMP12}.
Moreover, in \cite{MMP12} it was reformulated in matrix model terms,
what implies applicability of topological recursion ideas \cite{AMM/EO} --
this seems to be the way, explaining emergence of
skein relations in simple situations from the hypercube formalism.
In fact the relations, found in \cite{MMP12}
(which are probably equivalent to the older MOY relations \cite{MOY}),
are not quite recursive (what is also the case for the skein relations) --
still, in combination with enumerative methods, they are sufficient
to calculate fundamental HOMFLY for ordinary knots (this is currently checked
up to 10 intersections -- for the whole Rolfsen table).
However, application of the same computer  program (available at \cite{knotebook})
to virtual knots does {\it not} fix polynomials unambiguously.
It is the purpose of this paper to describe this (empirical) situation in some detail.

\bigskip

In the formalism of \cite{MMP12} the basic objects are ``quantum dimensions'',
associated with fat graphs, and above-mentioned relations are associated
with graph reshufflings.
What happens is that dimensions for some of the graphs remain undetermined --
but this never happens to graphs, appearing in consideration of ordinary
(non-virtual) knots and links.
A natural hypothesis is that we actually encounter {\it new} topological invariants,
which are not observable in the world of ordinary knots.
We did some checks ensuring that the new parameters are indeed
topological invariants -- but no proof is yet available.

One could suspect that such additional invariants exist -- already because of the
old puzzle of Kishino virtual knot \cite{Kishino,Kvirt},
which is a composite of two unknots,
not reducible to unknot by the Reidemeister-Kauffman moves.
While Kishino knot polynomial does not depend on any of our new parameters
-- and thus remains undistinguished from unknot by fundamental HOMFLY
-- new parameters do appear for a closely related knot.
This knot has the same shape of a planar diagram but all intersections
black, so it is a composite of two virtual trefoils.
Not only does its HOMFLY contain new parameters, it also contains
\textit{more} new parameters, then the virtual trefoil itself.

The paper is organized as follows. In Section \ref{sec:skein-method} we
remind how to calculate HOMFLY polynomial with help of skein relations.
In Section \ref{sec:skein-ambiguity} we show, that if we naively
apply this method to virtual knots, we are faced with ambiguities.
Then in Sections \ref{sec:hypercube-formalism} and \ref{sec:moy-rels}
we develop a proper language to address these issues
-- the hypercube formalism.
In Section \ref{sec:different-ways} we restate the ambiguity in this language,
where it takes the form of new parameters -- free values of some quantum dimensions.
In Section \ref{sec:computer-program} we describe an algorithm,
that can check, to what extent keeping these parameters free is consistent with topological invariance.
In Section \ref{sec:results} we present the outcome of this algorithm's work,
so impatient reader may immediately look here.
In Section \ref{sec:factorization-of-composites} we study, whether the ambiguity
can be constrained by imposing factorization of HOMFLY polynomial for composite knots.
We finish the main body of the paper by formulating two \textit{conjectures} about these
new empirically observed free parameters.
In Appendix \ref{sec:kishino-relatives} we consider in detail example of Kishino knot
-- showing that new parameters break factorization of composites.
Appendix \ref{sec:programs-work} contains more statistics on the results
obtained by algorithm of section \ref{sec:computer-program}.
Finally, in Appendix \ref{Appsymm} we show an alternative line of development of the hypercube formalism
-- using symmetric representation $[2]$ instead of antisymmetric $[1,1]$.

\section{Skein method = skein relations + Reidemeister moves}
\label{sec:skein-method}

One of the ways to calculate fundamental HOMFLY polynomial for a given planar diagram of a knot
is to use the skein relation

\begin{equation} \label{eq:skein-relation}
  A
  \begin{picture}(25,10)(-5,7)
    \put(0,0){\vector(1,1){20}}
    \put(20,0){\vector(-1,1){20}}
    \put(10,10){\circle{4}}
  \end{picture}
  - A^{-1}
  \begin{picture}(25,10)(-5,7)
    \put(0,0){\vector(1,1){20}}
    \put(20,0){\vector(-1,1){20}}
    \put(10,10){\circle*{4}}
  \end{picture}
  \ = \ z\
  \begin{picture}(25,10)(-5,7)
    \qbezier(0,0)(10,10)(0,20)
    \put(20,0){\qbezier(0,0)(-10,10)(0,20)}
    \put(0,20){\vector(-1,1){0}}
    \put(20,20){\vector(1,1){0}}
  \end{picture}
\end{equation}

\bigskip

\noindent
where $A=q^N$ and $z=q-q^{-1}$.
From the point of view of RT approach this relation is the property
of quantum ${\cal R}$-matrix in the fundamental representation
\be
\left({\cal R} - \frac{1}{Aq}\right)\left({\cal R} +\frac{q}{A}\right)=0
\ee
(for higher representations ${\cal R}$-matrix has more different eigenvalues,
and the story gets more involved).

The method is to gradually express HOMFLY
polynomial for a diagram through HOMFLY polynomials of simpler diagrams,
eventually expressing everything though completely
unknotted diagrams.

Each simplification step  consists of two substeps:

\begin{itemize}
\item skein relation is applied to some crossings of a planar diagram
\item some of resulting summands are simplified with the help of Reidemeister moves
\end{itemize}

Crucial is the second substep, without it the procedure would not terminate at all,
because one of the summands after the application of relation always has \textit{the same}
number of crossings as in the original diagram.
But can we always simplify this summand with help of Reidemeister moves,
such that it contains less crossings, than the diagram we started with?

The answer is yes -- if we choose crossings, at which skein relations are applied, in a clever way.
Namely, for non-virtual knots it is easy to observe that

\begin{prop}\label{pr:unknotting}
Every planar diagram can be unknotted by changing some of its crossings for inverse ones.
\end{prop}

\noindent
Thus, if we choose to apply skein relation precisely at all these ``unknotting'' crossings,
then the problematic summand with non-decreased number of crossings can definitely be simplified
-- even completely unknotted! --
with the help of Reidemeister moves -- because it {\it is} the unknot.
Thus {\it in principle} the procedure {\it can} calculate arbitrary fundamental HOMFLY.

The problem in practice is to recognize, which crossings are the unknotting ones?
If we are not concerned with efficiency and just want to show that algorithm terminates
in finite time,
we can choose the most naive, brute-force enumerative approach.

We just loop over all $2^{\# \text{crossings}}$ possible choices
and check, whether diagram we obtain by changing these crossings is an unknot.
To check, whether some diagram is an unknot is a famous ``unknotting problem''.
Again we can choose the most naive brute-force approach to it.

The number of Reidemeister moves, needed to unknot any planar diagram, is at most exponential
in the number of crossings \cite{exponential-unknotting}. Hence, the full search over
all possible sequences of Reidemeister moves (of restricted length) will take at most
doubly exponential time -- but \textit{finite} nonetheless. Hence, \textit{skein method will give
an answer for any planar diagram in finite time as well}.

Of course, in practice, it is possible to optimize this brute-force procedure in many ways.
First of all, it is very often possible to apply skein relation at just one crossing,
but in such a way, that the number of crossings of the problematic summand can be immediately
reduced with help of the second Reidemeister move. Second, there is a lot of smarter approaches
to the unknotting problem, which are at most exponential in the number of crossings
(though it is still unknown, whether there is a polynomial algorithm for that).

\paragraph{Example: trefoil}
Let's apply the skein method to get HOMFLY polynomial of a trefoil

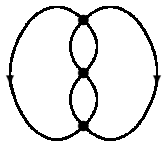
\begin{picture}(300,100)(-300,-40)
%
%
% \put(-90,0){\circle{30}}\put(-68.5,0){\circle{30}}
  \put(-80,0){
    \put(0,-10){\circle*{4}}\put(0,30){\circle*{4}}\put(0,10){\circle*{4}}
    \linethickness{0.2mm}
    \qbezier(0,-10)(10,0)(0,10) \qbezier(0,-10)(-10,0)(0,10)
    \put(0,20){\qbezier(0,-10)(10,0)(0,10) \qbezier(0,-10)(-10,0)(0,10)}
    % \put(0,10){\oval(40,40)}
    \qbezier(0,-10)(10,-20)(20,-10)
    \put(-20,0){\qbezier(0,-10)(10,-20)(20,-10)}
    \put(0,30){\qbezier(0,0)(10,10)(20,0)}
    \put(-20,30){\qbezier(0,0)(10,10)(20,0)}
    \put(-20,-10){\qbezier(0,0)(-15,20)(0,40)}
    \put(20,-10){\qbezier(0,0)(15,20)(0,40)}
    \put(-27.5,7){\vector(0,-1){2}}\put(27.5,7){\vector(0,-1){2}}
  }
%% \put(-84,-32){\mbox{${\cal L}$}}
%
\end{picture}

\noindent
First we substitute the upper black vertex by the white one, using skein relation,
to get

\begin{equation}
\begin{picture}(70,30)(-95,10)
%
%%
% \put(-90,0){\circle{30}}\put(-68.5,0){\circle{30}}
  \put(-80,0){
    \put(0,-10){\circle*{4}}\put(0,30){\circle*{4}}\put(0,10){\circle*{4}}
    \linethickness{0.2mm}
    \qbezier(0,-10)(10,0)(0,10) \qbezier(0,-10)(-10,0)(0,10)
    \put(0,20){\qbezier(0,-10)(10,0)(0,10) \qbezier(0,-10)(-10,0)(0,10)}
    % \put(0,10){\oval(40,40)}
    \qbezier(0,-10)(10,-20)(20,-10)
    \put(-20,0){\qbezier(0,-10)(10,-20)(20,-10)}
    \put(0,30){\qbezier(0,0)(10,10)(20,0)}
    \put(-20,30){\qbezier(0,0)(10,10)(20,0)}
    \put(-20,-10){\qbezier(0,0)(-15,20)(0,40)}
    \put(20,-10){\qbezier(0,0)(15,20)(0,40)}
    \put(-27.5,7){\vector(0,-1){2}}\put(27.5,7){\vector(0,-1){2}}
  }
\end{picture}
\ = \ \ A^2
\begin{picture}(70,30)(-115,10)
%
%
% \put(-90,0){\circle{30}}\put(-68.5,0){\circle{30}}
  \put(-80,0){
    \put(0,-10){\circle*{4}}\put(0,30){\circle{4}}\put(0,10){\circle*{4}}
    \linethickness{0.2mm}
    \qbezier(0,-10)(10,0)(0,10) \qbezier(0,-10)(-10,0)(0,10)
    \put(0,20){\qbezier(0,-10)(10,0)(0,10) \qbezier(0,-10)(-10,0)(0,10)}
    % \put(0,10){\oval(40,40)}
    \qbezier(0,-10)(10,-20)(20,-10)
    \put(-20,0){\qbezier(0,-10)(10,-20)(20,-10)}
    \put(0,30){\qbezier(0,0)(10,10)(20,0)}
    \put(-20,30){\qbezier(0,0)(10,10)(20,0)}
    \put(-20,-10){\qbezier(0,0)(-15,20)(0,40)}
    \put(20,-10){\qbezier(0,0)(15,20)(0,40)}
    \put(-27.5,7){\vector(0,-1){2}}\put(27.5,7){\vector(0,-1){2}}
  }
%% \put(-84,-32){\mbox{${\cal L}$}}
%
\end{picture}
- A z
\begin{picture}(50,30)(-115,10)
%
%
% \put(-90,0){\circle{30}}\put(-68.5,0){\circle{30}}
  \put(-80,0){
    \put(0,-10){\circle*{4}}\put(0,10){\circle*{4}}
    \linethickness{0.2mm}
    \qbezier(0,-10)(10,0)(0,10) \qbezier(0,-10)(-10,0)(0,10)
    \put(0,20){\qbezier(0,-10)(10,0)(5,10) \qbezier(0,-10)(-10,0)(-5,10)}
    \put(0,30){\qbezier(5,0)(10,10)(20,0)}
    \put(-20,30){\qbezier(0,0)(10,10)(15,0)}

    \qbezier(0,-10)(10,-20)(20,-10)
    \put(-20,0){\qbezier(0,-10)(10,-20)(20,-10)}
    \put(-20,-10){\qbezier(0,0)(-15,20)(0,40)}
    \put(20,-10){\qbezier(0,0)(15,20)(0,40)}
    \put(-27.5,7){\vector(0,-1){2}}\put(27.5,7){\vector(0,-1){2}}
  }
%% \put(-84,-32){\mbox{${\cal L}$}}
%
\end{picture}
\end{equation}

\bigskip

\bigskip

\bigskip

\noindent
We see, that the diagram in the first item still has the same number of crossings -- 3.
But we can apply the sequence of 2nd and 1st Reidemeister moves to unknot it -- thus it equals
$[N] =\frac{q^N-q^{-N}}{q-q^{-1}}=\frac{A-A^{-1}}{q-q^{-1}}$,
the quantum dimension of the fundamental representation,
which is usually associated with the unknot (this convention defines "non-normalized" HOMFLY,
if one instead associates just unity with the unknot, then all polynomials should be divided
by $[N]$ to provide "normalized" HOMFLY).
The diagram in the second summand is a Hopf link and has 2 crossings, so it's simpler,
than the original one -- we can say, that we've calculated it on the previous step
of the recursion, and it's equal to $A^2 [N]^2 - A z [N]$.
So the full answer is
\begin{equation}
  H_{\text{trefoil}} = A^2[N] - A z \left (A^2 [N]^2 - A z [N] \right )
  =   \left ( - A^4 + A^2 z^2  + 2 A^2 \right )\cdot [N]
\end{equation}

Though ``unknotting'' proposition \ref{pr:unknotting} is trivial for usual knots,
this statement often fails for virtual knots, as we shall shortly see in the next section.
Therefore, if we apply skein method to virtual knots, we can not always get a number
(i.e. a Laurent polynomial in $A$ and $q$) -- sometimes some un-simplifiable planar diagrams
remain in the answer. This, however, is not a bad thing: from a proper point of view
(hypercube formalism) these un-reduceable pictures are free parameters, that do not conflict
with topological invariance.
Hence, they can be treated as new parameters and HOMFLY polynomial
just starts to depend on a larger (very likely, infinite) set of variables -- and
still remains topological invariant.

\section{Skein method does not completely determine HOMFLY for virtual knots: example of Hopf link}
\label{sec:skein-ambiguity}

Let us now apply the skein method to the simplest virtual planar diagram -- the virtual Hopf link

\bigskip

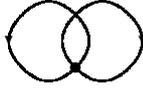
\begin{picture}(100,40)(-300,-20)
% \put(-90,0){\circle{30}}\put(-68.5,0){\circle{30}}
  \put(-80,0){
    \put(0,-10){\circle*{4}}
    \linethickness{0.2mm}
    \qbezier(0,-10)(10,0)(0,10) \qbezier(0,-10)(-10,0)(0,10)
    % \put(0,20){\qbezier(0,-10)(10,0)(0,10) \qbezier(0,-10)(-10,0)(0,10)}
    % \put(0,10){\oval(40,40)}
    \qbezier(0,-10)(10,-20)(20,-10)
    \put(-20,0){\qbezier(0,-10)(10,-20)(20,-10)}
    \put(0,10){\qbezier(0,0)(10,10)(20,0)}
    \put(-20,10){\qbezier(0,0)(10,10)(20,0)}
    \put(-20,-10){\qbezier(0,0)(-10,10)(0,20)}
    \put(20,-10){\qbezier(0,0)(10,10)(0,20)}
    \put(-25,0){\vector(0,-1){2}}\put(25,0){\vector(0,-1){2}}
  }
%% \put(-84,-32){\mbox{${\cal L}$}}
%
\end{picture}

\noindent
We immediately get into a trouble: the skein relation

\begin{equation}
  A^{-1}\
\begin{picture}(70,20)(-110,0)
% \put(-90,0){\circle{30}}\put(-68.5,0){\circle{30}}
  \put(-80,0){
    \put(0,-10){\circle*{4}}
    \linethickness{0.2mm}
    \qbezier(0,-10)(10,0)(0,10) \qbezier(0,-10)(-10,0)(0,10)
    % \put(0,20){\qbezier(0,-10)(10,0)(0,10) \qbezier(0,-10)(-10,0)(0,10)}
    % \put(0,10){\oval(40,40)}
    \qbezier(0,-10)(10,-20)(20,-10)
    \put(-20,0){\qbezier(0,-10)(10,-20)(20,-10)}
    \put(0,10){\qbezier(0,0)(10,10)(20,0)}
    \put(-20,10){\qbezier(0,0)(10,10)(20,0)}
    \put(-20,-10){\qbezier(0,0)(-10,10)(0,20)}
    \put(20,-10){\qbezier(0,0)(10,10)(0,20)}
    \put(-25,0){\vector(0,-1){2}}\put(25,0){\vector(0,-1){2}}
  }
%% \put(-84,-32){\mbox{${\cal L}$}}
%
\end{picture}
= A\
\begin{picture}(70,20)(-110,0)
% \put(-90,0){\circle{30}}\put(-68.5,0){\circle{30}}
  \put(-80,0){
    \put(0,-10){\circle{4}}
    \linethickness{0.2mm}
    \qbezier(0,-10)(10,0)(0,10) \qbezier(0,-10)(-10,0)(0,10)
    % \put(0,20){\qbezier(0,-10)(10,0)(0,10) \qbezier(0,-10)(-10,0)(0,10)}
    % \put(0,10){\oval(40,40)}
    \qbezier(0,-10)(10,-20)(20,-10)
    \put(-20,0){\qbezier(0,-10)(10,-20)(20,-10)}
    \put(0,10){\qbezier(0,0)(10,10)(20,0)}
    \put(-20,10){\qbezier(0,0)(10,10)(20,0)}
    \put(-20,-10){\qbezier(0,0)(-10,10)(0,20)}
    \put(20,-10){\qbezier(0,0)(10,10)(0,20)}
    \put(-25,0){\vector(0,-1){2}}\put(25,0){\vector(0,-1){2}}
  }
%% \put(-84,-32){\mbox{${\cal L}$}}
%
\end{picture}
- z \ \
\begin{picture}(70,20)(-110,0)
  \put(-80,0){
    \linethickness{0.2mm}
    \put(-20,0){\qbezier(0,-10)(10,-20)(15,-10)}
    \qbezier(5,-10)(10,-20)(20,-10)
    \qbezier(5,-10)(10,0)(0,10)
    \qbezier(-5,-10)(-10,0)(0,10)
    \put(0,10){\qbezier(0,0)(10,10)(20,0)}
    \put(-20,10){\qbezier(0,0)(10,10)(20,0)}
    \put(-20,-10){\qbezier(0,0)(-10,10)(0,20)}
    \put(20,-10){\qbezier(0,0)(10,10)(0,20)}
    \put(-25,0){\vector(0,-1){2}}\put(25,0){\vector(0,-1){2}}
  }
\end{picture}
\end{equation}

\bigskip

\bigskip

\noindent
does {\it not} express HOMFLY for virtual Hopf through HOMFLY for simpler knots!
Instead one of the items at the right hand side is mirror reflection of original one
at the l.h.s.
Even assuming that they are related by the symmetry $q \rightarrow  1/q$
(that sends $A \rightarrow A^{-1}$ and $z \rightarrow - z$),
we conclude, that the most what we can do
is to express part of HOMFLY for virtual Hopf as a number

\begin{equation}
  A^{-1}\
\begin{picture}(55,20)(-110,0)
% \put(-90,0){\circle{30}}\put(-68.5,0){\circle{30}}
  \put(-80,0){
    \put(0,-10){\circle*{4}}
    \linethickness{0.2mm}
    \qbezier(0,-10)(10,0)(0,10) \qbezier(0,-10)(-10,0)(0,10)
    % \put(0,20){\qbezier(0,-10)(10,0)(0,10) \qbezier(0,-10)(-10,0)(0,10)}
    % \put(0,10){\oval(40,40)}
    \qbezier(0,-10)(10,-20)(20,-10)
    \put(-20,0){\qbezier(0,-10)(10,-20)(20,-10)}
    \put(0,10){\qbezier(0,0)(10,10)(20,0)}
    \put(-20,10){\qbezier(0,0)(10,10)(20,0)}
    \put(-20,-10){\qbezier(0,0)(-10,10)(0,20)}
    \put(20,-10){\qbezier(0,0)(10,10)(0,20)}
    \put(-25,0){\vector(0,-1){2}}\put(25,0){\vector(0,-1){2}}
    \put(23,-15){\mbox{$_q$}}
  }
%% \put(-84,-32){\mbox{${\cal L}$}}
%
\end{picture}
\ \ - A\ \
\begin{picture}(55,20)(-110,0)
% \put(-90,0){\circle{30}}\put(-68.5,0){\circle{30}}
  \put(-80,0){
    \put(0,-10){\circle*{4}}
    \linethickness{0.2mm}
    \qbezier(0,-10)(10,0)(0,10) \qbezier(0,-10)(-10,0)(0,10)
    % \put(0,20){\qbezier(0,-10)(10,0)(0,10) \qbezier(0,-10)(-10,0)(0,10)}
    % \put(0,10){\oval(40,40)}
    \qbezier(0,-10)(10,-20)(20,-10)
    \put(-20,0){\qbezier(0,-10)(10,-20)(20,-10)}
    \put(0,10){\qbezier(0,0)(10,10)(20,0)}
    \put(-20,10){\qbezier(0,0)(10,10)(20,0)}
    \put(-20,-10){\qbezier(0,0)(-10,10)(0,20)}
    \put(20,-10){\qbezier(0,0)(10,10)(0,20)}
    \put(-25,0){\vector(0,-1){2}}\put(25,0){\vector(0,-1){2}}
    \put(25,-15){\mbox{$\frac{1}{q}$}}
  }
%% \put(-84,-32){\mbox{${\cal L}$}}
%
\end{picture}
\ =\  z [N]
\end{equation}

\bigskip

\noindent
while another linear combination
\begin{equation} \nonumber
\!\!\!\!\!\!\!\!\!\!\!\!\!\!\!\!\!\!\!\!\!
  A^{-1}\
\begin{picture}(55,20)(-110,0)
% \put(-90,0){\circle{30}}\put(-68.5,0){\circle{30}}
  \put(-80,0){
    \put(0,-10){\circle*{4}}
    \linethickness{0.2mm}
    \qbezier(0,-10)(10,0)(0,10) \qbezier(0,-10)(-10,0)(0,10)
    % \put(0,20){\qbezier(0,-10)(10,0)(0,10) \qbezier(0,-10)(-10,0)(0,10)}
    % \put(0,10){\oval(40,40)}
    \qbezier(0,-10)(10,-20)(20,-10)
    \put(-20,0){\qbezier(0,-10)(10,-20)(20,-10)}
    \put(0,10){\qbezier(0,0)(10,10)(20,0)}
    \put(-20,10){\qbezier(0,0)(10,10)(20,0)}
    \put(-20,-10){\qbezier(0,0)(-10,10)(0,20)}
    \put(20,-10){\qbezier(0,0)(10,10)(0,20)}
    \put(-25,0){\vector(0,-1){2}}\put(25,0){\vector(0,-1){2}}
    \put(23,-15){\mbox{$_q$}}
  }
%% \put(-84,-32){\mbox{${\cal L}$}}
%
\end{picture}
+ A \
\begin{picture}(55,20)(-110,0)
% \put(-90,0){\circle{30}}\put(-68.5,0){\circle{30}}
  \put(-80,0){
    \put(0,-10){\circle*{4}}
    \linethickness{0.2mm}
    \qbezier(0,-10)(10,0)(0,10) \qbezier(0,-10)(-10,0)(0,10)
    % \put(0,20){\qbezier(0,-10)(10,0)(0,10) \qbezier(0,-10)(-10,0)(0,10)}
    % \put(0,10){\oval(40,40)}
    \qbezier(0,-10)(10,-20)(20,-10)
    \put(-20,0){\qbezier(0,-10)(10,-20)(20,-10)}
    \put(0,10){\qbezier(0,0)(10,10)(20,0)}
    \put(-20,10){\qbezier(0,0)(10,10)(20,0)}
    \put(-20,-10){\qbezier(0,0)(-10,10)(0,20)}
    \put(20,-10){\qbezier(0,0)(10,10)(0,20)}
    \put(-25,0){\vector(0,-1){2}}\put(25,0){\vector(0,-1){2}}
    \put(25,-15){\mbox{$\frac{1}{q}$}}
  }
%% \put(-84,-32){\mbox{${\cal L}$}}
%
\end{picture}
\end{equation}

\bigskip

\noindent
remains unconstrained by skein relation.

\bigskip

Of course, one may wonder, whether it is possible to first twist the planar diagram
of a virtual Hopf link in some way, and only then apply skein method, such that in
the end we would still get a number.
This question is surprisingly hard to answer and is, in a sense, the central question
of this paper.
The best we can do now is to present checks, that such a twisting is {\it not} possible,
when we consider diagrams with up to and including 7 crossings.
To address the issue systematically, we switch
in the next section to the hypercube formalism .

\section{Hypercube formula: solution to skein relation}
\label{sec:hypercube-formalism}

In \cite{MMP12} we wrote the hypercube-style formula, that expressed HOMFLY polynomial
for a virtual knot as a sum over amputations of edges of its \textit{dessin d'enfant}
\footnote{
The term ``\textit{dessin d'enfant}'' or ``\textit{dessin}'', for short,
means exactly the same as ``fat graph'' or ``ribbon graph''
and in the following we use all these terms interchangeably.}.

\newcommand\GLC[0]{\Gamma^{{\cal L}_c}}
\be \label{eq:hypercube-formula}
\  H^{{\cal L}_c}_{_\Box}
=   q^{(N-1)\left( n_{\bullet}(\GLC)-n_{\circ}(\GLC)\right)}
\sum_{\gamma\subseteq \GLC} (-q)^{n_{\bullet}(\gamma) - n_\circ(\gamma)}
\cdot D_\gamma(q,N)
\phantom{5^{5^{5^{5^{5^5}}}}}\!\!\!\!\!\!\!\!\!\!\!\!
\ee
%% \begin{verbatim}
%%  repeat the summation formula to be self-contained
%% \end{verbatim}

Here $\GLC$ is the fat graph, that is associated to the planar diagram $\mathcal{L}_c$.
Vertices of this fat graph are Seifert cycles of a planar diagram and edges are its crossings
(in \cite{MMP12} everything is explained in more detail, with examples).
$n_\bullet$ and $n_\circ$ are number of black and white edges, respectively.
Quantum dimensions $D_\gamma$ are functions of $q$ and $N$, that are associated to any
fat graph. The sum in the formula runs over all subgraphs $\gamma$ of $\GLC$, such that
they contain all the vertices of $\GLC$, but may not contain some edges (hence the name ``sum over amputations'').

In fact, this formula almost appears already in the old work   \cite{MOY}
-- but for planar fat graphs only.
In this case the formula can be  easily explained from RT point of view.
Namely, if one
takes expressions for fundamental ${\cal R}$-matrix and its inverse through the projector
onto antisymmetric representation $P_{[1,1]}$
%% \begin{verbatim}
%% the actual expressions go here
%% \end{verbatim}
\be
{\cal R} = q^{-N}\cdot\left(\frac{1}{q}\cdot I \otimes I - [2] P_{[1,1]}\right) \\ \nn
{\cal R}^{-1} = q^N\cdot\left(q\cdot I \otimes I - [2] P_{[1,1]}\right)
\ee
then the hypercube formula \eqref{eq:hypercube-formula},
is just the result of expanding all the brackets in the RT tensor contraction.
The $q$-dimensions $D_\gamma$ in the usual RT-language are called ``suitably defined quantum traces''.
The authors of \cite{MOY} provide one of possible consistent definitions
of these traces and then derive from this definition some properties.
In particular, they show that their definition satisfies recursion relations (MOY-relations),
that hypercube formula \eqref{eq:hypercube-formula} is invariant w.r.t all Reidemeister moves
and that it satisfies skein relation \eqref{eq:skein-relation}.

In the rest of this section and in the next one we strengthen these statements a little bit,
in particular, make them applicable also to virtual knots.

\bigskip

First, observe, that
\begin{prop}
  Hypercube formula \eqref{eq:hypercube-formula} satisfies skein relation \eqref{eq:skein-relation}
  regardless of what $D_\gamma$'s are.
\end{prop}

%% As we shall see later, these properties can be implied from topological invariance
%% of \eqref{eq:hypercube-formula}, \textit{without any explicit definition of $D_\gamma$}.

%% Now observe that formula \eqref{eq:hypercube-formula} actually satisfies skein relations,
%% regardless of what $D_\gamma$'s are equal to.

Indeed, consider some black vertex. All summands in \eqref{eq:hypercube-formula}
split into two big groups: the ones, where the edge, corresponding to this vertex, is kept,
and the ones, where it is amputated. Hence, we can write (with  self-explanatory notation)
\begin{equation}
  H_\bullet = q^{N-1} (-q) H_{\bullet\ \text{kept}} + q^{N-1} H_{\bullet\ \text{amp}}
\end{equation}
where  $q$-factors, produced by this vertex, are explicitly written down --
and thus neither $H_{\bullet\ \text{kept}}$, nor $H_{\bullet\ \text{amp}}$
contain any factors, dependent on the color of our selected vertex.

Now, if this vertex is changed for white one, we can write
\begin{equation}
  H_\circ = q^{1 - N} (-q^{-1}) H_{\circ\ \text{kept}} + q^{1-N} H_{\circ\ \text{amp}}
\end{equation}

Since all color dependence on the chosen vertex is
absent in $H_{\bullet\ \text{kept}}$, $H_{\circ\ \text{kept}}$,
$H_{\bullet\ \text{amp}}$ and $H_{\circ\ \text{amp}}$,
we actually have $H_{\bullet\ \text{kept}} = H_{\circ\ \text{kept}}$ and
$H_{\bullet\ \text{amp}} = H_{\circ\ \text{amp}}$. \
This allows us to write
\begin{equation}
  A^{-1} H_\bullet - A H_\circ = (q^{-1} - q)\cdot H_{amp} = (q^{-1} - q)\cdot H_{||} = - z H_{||},
\end{equation}
which is exactly the skein relation \eqref{eq:skein-relation}.

\paragraph{Example: virtual Hopf link}
If we apply the hypercube formula to the virtual Hopf link above, we get
%% \begin{verbatim}
%% the expression for HOMFLY for vHopf through 1-vertex 1-edge loop
%% \end{verbatim}

\newcommand\Dloop[0]{
D_{
  \begin{picture}(10,10)(-5,-5)
    \put(0,0){\circle{10}}
    \put(-5,0){\circle*{4}}
  \end{picture}
}
}

\begin{equation}
H_{\text{vHopf}} = q^{N-1} \left ( D_{\bullet} - q \Dloop \right )
\end{equation}

While $D_\bullet$ poses no questions -- it is just a familiar dimension $[N]$ of the unknot,
we actually see the new unknown parameter: value of $q$-dimension for the {\it dessin}
with one vertex and one loop-edge $\Dloop$.
Furthermore, we see that this answer satisfies skein relation,
in accordance with our general argument -- regardless of what we substitute for $\Dloop$.

%% \begin{verbatim}
%% illustration, that our answer indeed satisfies skein
%% \end{verbatim}

\begin{equation}
H_{\text{vHopf} \bullet \rightarrow \circ} = q^{1 - N} \left ( D_{\bullet} - q^{-1} \Dloop \right )
\end{equation}

\begin{equation}
H_{\text{vHopf} \bullet \rightarrow ||} = D_{\bullet}
\end{equation}

\begin{equation}
  A^{-1}\cdot H_{\text{vHopf}} - A\cdot H_{\text{vHopf} \bullet \rightarrow \circ}
  = (q^{-1} - q)\cdot H_{\text{vHopf} \bullet \rightarrow ||}
\end{equation}

\bigskip

Thus, we see, that out of two constituents of the skein method -- skein relation and Reidemeister moves,--
the hypercube formula \eqref{eq:hypercube-formula} is sensitive only to the first one.
It is quite natural to expect that it is the second constituent -- Reidemeister invariance --
which puts some constraints on the so far completely free parameters $D_\gamma$.
As we will see in the following section, this is indeed the case -- but
in the case of virtual knots these constraints are not sufficient to fully determine all the $D_\gamma$.

\bigskip

We finish this section with a small remark:

\begin{prop}
{Hypercube formula is not the most general solution to skein relations}
\end{prop}

Though formula \eqref{eq:hypercube-formula}, with arbitrary $D_\gamma$,
solves skein relation \eqref{eq:skein-relation}, it still has drawbacks.
For instance, it automatically respects all the virtual Reidemeister moves \cite{MMP12}.
Moreover, it also respects the following additional move
-- interchange of the usual and virtual crossings, which is not in the list \cite{MMP12}.
%% \begin{verbatim}
%% the move that changes usual and virtual intersections
%% \end{verbatim}
\begin{equation}
  \begin{picture}(20, 20)(0,0)
    \qbezier(-10,-20)(10,0)(-10,20)
    \put(-10,0){\qbezier(10,-20)(-10,0)(10,20)}
    \put(-5.5,14){\circle*{4}}
  \end{picture}
  =
  \begin{picture}(20, 20)(-30,0)
    \qbezier(-10,-20)(10,0)(-10,20)
    \put(-10,0){\qbezier(10,-20)(-10,0)(10,20)}
    \put(-5.5,-14){\circle*{4}}
  \end{picture}
\end{equation}

\bigskip

\bigskip

\noindent
This indicates, that it could be possible to write a more relaxed formula,
that would be just the generic form of a solution to skein relation.
To invent such a formula is a very interesting open question, which,
however, is left out of the scope of this paper.

\section{Mild assumptions + Reidemeister moves = MOY-relations}\label{sec:moy-rels}

So, formula \eqref{eq:hypercube-formula} automatically incorporates skein relation.
But there is another component of skein method, namely, application of Reidemeister moves
to actually unknot the planar diagram after application of skein relations.
This Reidemeister invariance would now put some constraints on components
of the formula \eqref{eq:hypercube-formula}, namely, on dimensions $D_\gamma$.

In fact, theorem 5.1 of \cite{MMP12} can be reinterpreted in the following way:
suppose dimensions $D_\gamma$ satisfy
\begin{itemize}
\item factorization property: $D_\gamma = D_{\gamma_1}\cdot D_{\gamma_2} \ \ \ \ \
\text{for} \ \ \ \gamma = \gamma_1\cup \gamma_2
\ \ \text{ with} \ \ \gamma_1\cap\gamma_2 = \emptyset$
\item normalization: dimension of an isolated vertex is $[N]$, $D_{\bullet} = [N]$
\end{itemize}

\noindent
Then invariance of \eqref{eq:hypercube-formula} w.r.t Reidemeister moves
implies that $D_\gamma$ are ``local'' functions of fat-graph $\gamma$
(as Reidemeister moves themselves can be applied to any small part of planar diagram)
-- they
change nicely under certain transformations of the graph, which change only the small
piece of it.

These transformations are, in fact, well-known in the literature: they are
celebrated MOY-relations \cite{MOY}. In particular, it is these relations, that
inspired Khovanov and Rozansky to invent their categorification of HOMFLY polynomials
\cite{KR}.
However, in \cite{MOY} these relations are corollaries (Lemmas 2.2, 2.3, 2.4 and 2.5)
of an independent (non-recursive) definition of $D_\gamma$.
We show (\cite{MMP12}) that for \textit{any} definition of $D_\gamma$
(satisfying factorization and normalization properties) these
relations should be satisfied, to ensure topological invariance

%% \begin{verbatim}
%%   N-rules, again to be self-contained
%% \end{verbatim}
\be \label{eq:n-1-rule} \label{eq:moy-rels-b}
&  \begin{picture}(300,50)(0,-10)
\put(100,0){
    \linethickness{0.4mm} \thicklines
    \qbezier(0,0)(20,20)(0,40)
    \put(0,40){\vector(-1,1){0}}
    \put(35,10){\vector(1,0){0}}
    \put(30,20){\circle{20}}
        {\linethickness{0.15mm} \put(10,20){\line(1,0){10}}}
        \put(50,17) {$\simeq\ \ [N-1]$}
        \put(110,0){\vector(0,1){40}}
}
  \end{picture}
\ee

\be \label{eq:2-rule}
& \begin{picture}(300,60)(-80,-20)
    \linethickness{0.4mm} \thicklines
    \qbezier(0,0)(20,20)(0,40)
    \put(0,40){\vector(-1,1){0}}
    \put(40,0){\qbezier(0,0)(-20,20)(0,40)}
    \put(40,40){\vector(1,1){0}}
        {\linethickness{0.15mm}
          \put(8,30){\line(1,0){24}}
          \put(8,10){\line(1,0){24}}
        }
        \put(50,17) {$\simeq\ \ [2]$}
        \put(80,0){
          \linethickness{0.4mm} \thicklines
          \qbezier(0,0)(20,20)(0,40)
          \put(0,40){\vector(-1,1){0}}
          \put(40,0){\qbezier(0,0)(-20,20)(0,40)}
          \put(40,40){\vector(1,1){0}}
              {\linethickness{0.15mm}
                \put(10,20){\line(1,0){20}}
              }
        }
  \end{picture}
\ee

\be \label{eq:n-2-rule}
& \begin{picture}(300,60)(-80,-20)
    \linethickness{0.4mm} \thicklines
    \put(-40,0){\qbezier(0,0)(20,20)(0,40) \put(0,40){\vector(-1,1){0}}}
    \put(40,0){\qbezier(0,0)(-20,20)(0,40)}
    \put(40,0){\vector(1,-1){0}}
    \put(-40,0){\linethickness{0.15mm}
      \put(10,20){\line(1,0){20}}
    }
    \put(0,0){\linethickness{0.15mm}
      \put(10,20){\line(1,0){20}}
    }
    \put(0,20){\circle{20} \put(5,10){\vector(1,0){0}} \put(-5,-10){\vector(-1,0){0}}}

    \put(50,17) {$\simeq\ \ [N-2]$}
    \put(100,0) {
      \linethickness{0.4mm} \thicklines
      \qbezier(0,0)(20,20)(0,40)
      \put(0,40){\vector(-1,1){0}}
      \put(40,0){\qbezier(0,0)(-20,20)(0,40)}
      \put(40,0){\vector(1,-1){0}}
    }
    \put(150,17){$+$}
    \put(165,0) {
      \linethickness{0.4mm} \thicklines
      \qbezier(0,0)(20,20)(40,0)
      \put(0,40){\vector(-1,1){0}}
      \put(40,40){\qbezier(0,0)(-20,-20)(-40,0)}
      \put(40,0){\vector(1,-1){0}}
    }
  \end{picture}
\ee

\be \label{eq:n-3-rule}
& \begin{picture}(300,50)(-10,-10)
    \put(-40,0) {
      \linethickness{0.4mm} \thicklines
      \qbezier(0,0)(0,20)(-20,20) \put(-20,20){\vector(-1,0){0}}
      \put(20,0){\qbezier(0,0)(0,20)(20,20) \put(0,0){\vector(0,-1){0}}}
      \put(20,40){\qbezier(0,0)(-10,-10)(-20,0) \put(0,0){\vector(2,1){0}}}
      \put(10,20){\circle{16}}
      \put(10,0){
        \linethickness{0.15mm} \thinlines
        \qbezier(-12,11)(-12,11)(-6,15)
      }
      \put(10,0){
        \linethickness{0.15mm} \thinlines
        \qbezier(12,11)(12,11)(6,15)
      }
      \put(10,28){
        \linethickness{0.15mm} \thinlines
        \qbezier(0,0)(0,5)(0,7)
      }
    }
    \put(10,17){$-\ [N-3]$}
    \put(80,0) {
      \linethickness{0.4mm} \thicklines
      \qbezier(0,0)(0,20)(-20,20) \put(-20,20){\vector(-1,0){0}}
      \put(20,0){\qbezier(0,0)(0,20)(20,20) \put(0,0){\vector(0,-1){0}}}
      \put(20,40){\qbezier(0,0)(-10,-10)(-20,0) \put(0,0){\vector(2,1){0}}}
    }
    \put(130,17){$\simeq$}
    \put(210,0) {
      \put(-40,0) {
        \linethickness{0.4mm} \thicklines
        \qbezier(0,40)(0,20)(-20,20) \put(-20,20){\vector(-1,0){0}}
        \put(20,0){\qbezier(0,40)(0,20)(20,20) \put(0,40){\vector(0,1){0}}}
        \put(20,0){\qbezier(0,0)(-10,10)(-20,0) \put(0,0){\vector(2,-1){0}}}
        \put(10,20){\circle{16}}
        \put(10,40){
          \linethickness{0.15mm} \thinlines
          \qbezier(-12,-11)(-12,-11)(-6,-15)
        }
        \put(10,40){
          \linethickness{0.15mm} \thinlines
          \qbezier(12,-11)(12,-11)(6,-15)
        }
        \put(10,5){
          \linethickness{0.15mm} \thinlines
          \qbezier(0,0)(0,5)(0,7)
        }
      }
      \put(10,17){$-\ [N-3]$}
      \put(80,0) {
        \linethickness{0.4mm} \thicklines
        \qbezier(0,40)(0,20)(-20,20) \put(-20,20){\vector(-1,0){0}}
        \put(20,0){\qbezier(0,40)(0,20)(20,20) \put(0,40){\vector(0,1){0}}}
        \put(20,0){\qbezier(0,0)(-10,10)(-20,0) \put(0,0){\vector(2,-1){0}}}
      }
    }
  \end{picture}
\ee

\be \label{eq:1-rule}  \label{eq:moy-rels-e}
& \begin{picture}(300,60)(-30,-10)
    \linethickness{0.4mm} \thicklines
    \put(0,0) {
      \linethickness{0.4mm} \thicklines
      \qbezier(0,0)(0,20)(0,40)
      \put(0,40){\vector(0,1){0}}
      %% \put(40,0){\qbezier(0,0)(-20,20)(0,40)}
      %% \put(40,0){\vector(1,-1){0}}
      \put(-10,10){\linethickness{0.15mm}
        \put(10,20){\line(1,0){20}}
      }
      \put(-10,-10){\linethickness{0.15mm}
        \put(10,20){\line(1,0){20}}
      }
    }
    \put(20,0) {
      \linethickness{0.4mm} \thicklines
      \qbezier(0,0)(0,20)(0,40)
      \put(0,40){\vector(0,1){0}}
      %% \put(40,0){\qbezier(0,0)(-20,20)(0,40)}
      %% \put(40,0){\vector(1,-1){0}}
      \put(-10,0){\linethickness{0.15mm}
        \put(10,20){\line(1,0){20}}
      }
    }
    \put(40,0) {
      \linethickness{0.4mm} \thicklines
      \qbezier(0,0)(0,20)(0,40)
      \put(0,40){\vector(0,1){0}}
    }
    \put(50,17){$-$}
    \put(65,0) {
      \linethickness{0.4mm} \thicklines
      \put(0,0) {
        \linethickness{0.4mm} \thicklines
        \qbezier(0,0)(0,20)(0,40)
        \put(0,40){\vector(0,1){0}}
        %% \put(40,0){\qbezier(0,0)(-20,20)(0,40)}
        %% \put(40,0){\vector(1,-1){0}}
        \put(-10,0){\linethickness{0.15mm}
          \put(10,20){\line(1,0){20}}
        }
      }
      \put(20,0) {
        \linethickness{0.4mm} \thicklines
        \qbezier(0,0)(0,20)(0,40)
        \put(0,40){\vector(0,1){0}}
      }
      \put(40,0) {
        \linethickness{0.4mm} \thicklines
        \qbezier(0,0)(0,20)(0,40)
        \put(0,40){\vector(0,1){0}}
      }
    }
    \put(120,17){$\simeq$}
    \put(140,0) {
      \linethickness{0.4mm} \thicklines
      \put(0,0) {
        \linethickness{0.4mm} \thicklines
        \qbezier(0,0)(0,20)(0,40)
        \put(0,40){\vector(0,1){0}}
        %% \put(40,0){\qbezier(0,0)(-20,20)(0,40)}
        %% \put(40,0){\vector(1,-1){0}}
        \put(-10,0){\linethickness{0.15mm}
          \put(10,20){\line(1,0){20}}
        }
      }
      \put(20,0) {
        \linethickness{0.4mm} \thicklines
        \qbezier(0,0)(0,20)(0,40)
        \put(0,40){\vector(0,1){0}}
        \put(-10,10){\linethickness{0.15mm}
          \put(10,20){\line(1,0){20}}
        }
        \put(-10,-10){\linethickness{0.15mm}
          \put(10,20){\line(1,0){20}}
        }
      }
      \put(40,0) {
        \linethickness{0.4mm} \thicklines
        \qbezier(0,0)(0,20)(0,40)
        \put(0,40){\vector(0,1){0}}
      }
      \put(50,17){$-$}
      \put(65,0) {
        \linethickness{0.4mm} \thicklines
        \put(0,0) {
          \linethickness{0.4mm} \thicklines
          \qbezier(0,0)(0,20)(0,40)
          \put(0,40){\vector(0,1){0}}
        }
        \put(20,0) {
          \linethickness{0.4mm} \thicklines
          \qbezier(0,0)(0,20)(0,40)
          \put(0,40){\vector(0,1){0}}
          \put(-10,0){\linethickness{0.15mm}
            \put(10,20){\line(1,0){20}}
          }
        }
        \put(40,0) {
          \linethickness{0.4mm} \thicklines
          \qbezier(0,0)(0,20)(0,40)
          \put(0,40){\vector(0,1){0}}
        }
      }
    }
  \end{picture}
\ee

Armed with hypercube formula \eqref{eq:hypercube-formula} and MOY-relations \eqref{eq:moy-rels-b}-\eqref{eq:moy-rels-e}
we are ready to address the question, posed by section \ref{sec:skein-ambiguity}.
First, we rephrase the question in the hypercube language.

Suppose, we calculate HOMFLY polynomial for some planar diagram with the help of hypercube method.
First we write down the hypercube formula \eqref{eq:hypercube-formula} and obtain HOMFLY as
a linear combination of dimensions for different fat graphs.
Then, using MOY-relations \eqref{eq:moy-rels-b}-\eqref{eq:moy-rels-e} we gradually simplify these fat
graphs, such that, hopefully, in the end they become collections of isolated vertices
and we get a Laurent polynomial in $z$ and $A$.

But is it always possible? Thanks to our analysis in section \ref{sec:skein-method}
we know, that for {\it dessins}, that come from planar diagrams of non-virtual knots there
is always at least one way to decompose them, such that result is a number.
Moreover, since MOY \cite{MOY} give explicit formula for these dimensions for planar
fat graphs, we know that any two different decomposition paths for a planar fat graph
would give the same answer.

But for non-planar {\it dessins} we have neither proposition \ref{pr:unknotting},
nor MOY-formula \cite{MOY}, therefore, whether

\begin{itemize}
\item for every non-planar {\it dessin} it is possible to obtain a \textit{numeric}
  (i.e., not containing other {\it dessins}) answer via application of MOY-relations;
\item different paths of decomposition of a given {\it dessin} would give \textit{same} answer
  as a polynomial of simpler {\it dessins} (i.e. that there are no additional relations
  between various $D_\gamma$'s, except MOY)
\end{itemize}

\noindent
are open questions -- and these are the proper hypercube reformulation
of the questions in section \ref{sec:skein-ambiguity}.

In the following we investigate these questions. Actually we do not have general theoretical arguments, and thus resort to computer checks
(which, even if very excessive and persuasive, are by no means sufficient).

\begin{rmk}
Note, that neither the flip-rule (\cite{MMP12}, eq.(24)) nor the $1/[N]$-factorization property
(\cite{MMP12}, eq.(11)) are part of MOY-relations, which are required by topological invariance.
They are additional properties, that are most naturally present, if one takes matrix-model
point of view, as inspired by \cite{DM3}).
However, an empirical claim of this paper is that they can be safely dropped
-- this will not break topological invariance of the answer.
What does this mean on the matrix-model side of the story is an interesting question
for future research.
\end{rmk}

\begin{rmk}
Also note, that relation \eqref{eq:n-3-rule} was not among the original MOY-relations \cite{MOY}.
This is because the 3rd Reidemeister move with antiparallel orientation of strands
(which actually implies \eqref{eq:n-3-rule}) is, topologically, a consequence
of parallel 3rd Reidemeister move and antiparallel 2nd Reidemeister move, so MOY decided to omit it.
This means that if we did the same, if we had not use \eqref{eq:n-3-rule} to simplify {\it dessins},
some {\it dessins} would seem indecomposable at first appearance.
But then, when considering some {\it dessins} with bigger number of edges,
we would suddenly see that they have at least two different decomposition paths:
one allows to completely simplify this bigger {\it dessin} to a number, while the
other expresses it though previously indecomposable smaller {\it dessin}.
Therefore, we would conclude, that this small indecomposable {\it dessin} should have some
definite numeric value. But, of course, since we are actually using \eqref{eq:n-3-rule}
in our search of decompositions, we get the numeric value for the small {\it dessin} straight away.
\end{rmk}

\section{Different ways to decompose {\it dessins}: tests for topological invariance}
\label{sec:different-ways}

%% \begin{verbatim}
%% (brief) description of the program
%% \end{verbatim}

The recursion relations \eqref{eq:moy-rels-b}-\eqref{eq:moy-rels-e} allow to express more complicated
{\it dessins}
(i.e. the ones with bigger number of edges) in terms of simpler ones.
However, often there is more than one possible way to do the decomposition.

The simplest example of this phenomenon is a cycle {\it dessin} with two vertices and two edges.
It can be simplified either with help of $[N-2]$-rule (\ref{eq:n-2-rule}) or $[2]$-rule (\ref{eq:2-rule}).

%% \begin{verbatim}
%% picture of two ways to decompose 2-vertex cycle
%% \end{verbatim}
\be \label{eq:example-simple-rel}
  \begin{picture}(20,20)
    \put(0,0){\circle*{4}}
    \qbezier(0,0)(10,10)(20,0)
    \qbezier(0,0)(10,-10)(20,0)
    \put(20,0){\circle*{4}}
  \end{picture}
  = [N-2]\ \  \begin{picture}(10,20) \put(0,0){\circle*{4}} \end{picture}
  + \ \ \begin{picture}(20,20) \put(0,0){\circle*{4}} \put(10,0){\circle*{4}} \end{picture}
  = [N] \left ( [N-2] + [N] \right )
  \\ \nn
  \begin{picture}(20,20)
    \put(0,0){\circle*{4}}
    \qbezier(0,0)(10,10)(20,0)
    \qbezier(0,0)(10,-10)(20,0)
    \put(20,0){\circle*{4}}
  \end{picture}
  = [2] \ \
  \begin{picture}(20,20)
    \put(0,0){\circle*{4}}
    \qbezier(0,0)(10,0)(20,0)
    \put(20,0){\circle*{4}}
  \end{picture}
  = [2] [N] [N-1]
\ee

\noindent Thanks to the identity between $q$-numbers
\begin{equation}
  [N] + [N-2] = [2][N-1]
\end{equation}
these two ways actually give the same answer. But is it always the case?
Do any two different ways to decompose some complicated {\it dessin} always give the same answer?

Clearly, for some {\it dessins} it is not possible to find a sequence of MOY-moves, that expresses
them through {\it dessins} with fewer edges, for example, {\it dessin} appearing in calculation of
HOMFLY for virtual Hopf link

\begin{equation} \nn
\begin{picture}(10,10)(-5,-5)
  \put(0,0){\circle{10}}
  \put(-5,0){\circle*{4}}
\end{picture}
\end{equation}

\noindent is such a {\it dessin}. We call such {\it dessins} \textit{\bf atomic}.

Continuing the analogy, {\it dessins}, that can be decomposed, for example
\begin{equation} \nn
  \begin{picture}(20,20)
    \put(0,0){\circle*{4}}
    \qbezier(0,0)(10,10)(20,0)
    \qbezier(0,0)(10,-10)(20,0)
    \put(20,0){\circle*{4}}
  \end{picture}
\end{equation}
we call \textit{\bf molecular}.

Now, consider some {\it dessin}. All the ways to apply [N-1]-, [N-2]- or [2]-rule
to it once should give the same answer (since we assume that value of every {\it dessin} is definite).
Hence, we require that results of application of there rules are equal.
This gives us some relations on {\it dessins} with fewer number of edges.
We call such relations \textit{\bf simple relations}.

For example, relation \eqref{eq:example-simple-rel} is an example of such relation.

Sometimes, it is also possible to apply [N-3]- or [1]-rule to a {\it dessin}.
This also gives some relation, but this time between {\it dessins} with same number of edges.
We call such relations \textit{\bf cluster relations}.

What we ultimately want to know, is whether these simple and cluster relations
give some non-trivial relations on atomic {\it dessins}, in particular, whether it follows
from them, that

\be \label{eq:flip-and-1-n-rels}
\begin{picture}(10,10)(-5,-5)
  \put(0,0){\circle{10}}
  \put(-5,0){\circle*{4}}
\end{picture}
= (-)\ \   \begin{picture}(20,20)(0,-3)
    \put(0,0){\circle*{4}}
    \qbezier(0,0)(10,0)(20,0)
    \put(20,0){\circle*{4}}
  \end{picture}
\ \ =\ \ - [N][N-1] \nn\\ \nn \\
\begin{picture}(10,10)(-5,-5)
  \put(0,0){\circle{10}}
  \put(-10,0){\circle{10}}
  \put(-5,0){\circle*{4}}
\end{picture}
= \frac{1}{[N]}
\begin{picture}(10,10)(-5,-5)
  \put(0,0){\circle{10}}
  \put(-5,0){\circle*{4}}
\end{picture}^2 = [N][N-1]^2
\ee

\bigskip

\noindent
The first equation is a particular case of flip-rule (\cite{MMP12}, eq.(24))
and the second is a particular case of $\frac{1}{[N]}$-factorization (\cite{MMP12}, eq.(11)).
We would like to know, whether these, or similar, formulas are corollaries of
MOY-relations \eqref{eq:moy-rels-b}-\eqref{eq:moy-rels-e}.

MOY-relations in the hypercube formalism directly correspond to the Reidemeister moves
in the skein method. Hence, when we study, which {\it dessins} are atomic, which are molecular
and what are simple and cluster relations between {\it dessins}, we actually learn, what
restrictions on possible choice of values of $D_\gamma$'s does the topological invariance impose.
This is useful, since \textbf{if an atomic {\it dessin} is not restricted by any relations,
  we can treat it as a new variable in a HOMFLY polynomial}.
Two candidates for such new variables are {\it dessins}
$\begin{picture}(10,10)(-5,-5)
  \put(0,0){\circle{10}}
  \put(-5,0){\circle*{4}}
\end{picture}$ and
$\begin{picture}(20,10)(-15,-5)
  \put(0,0){\circle{10}}
  \put(-10,0){\circle{10}}
  \put(-5,0){\circle*{4}}
\end{picture}$ (provided we wouldn't find any of relations \eqref{eq:flip-and-1-n-rels}).

The extensive search for these (and other) relations, however, is easier said than done. This is because
the number of essentially distinct {\it dessins} with given number of edges grows very fast
(faster, than Catalan numbers). Hence, a computer program is required.
We've written such a program and briefly outline here the logic of its work.

\section{Computer program}\label{sec:computer-program}

\begin{itemize}
\item The program works by layers -- at each layer it looks at {\it dessins} with given number of edges.
  It is assumed, that expressions for all {\it dessins} with fewer edges in terms of atomic {\it dessins} are already
  found.
\item It splits {\it dessins} on a given layer into clusters -- {\it dessins} that can be transformed into
  one another with help of [N-3]-rule and 1-rule (modulo {\it dessins} with fewer number of edges).
  For example, these two {\it dessins} on layer 3 are in the same cluster (via [N-3]-rule)
  \begin{equation} \nn
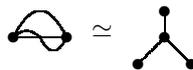
    \begin{picture}(20,20)
      \put(0,0){\circle*{4}}
      \put(20,0){\circle*{4}}
      \qbezier(0,0)(10,0)(20,0)
      \qbezier(0,0)(5,5)(10,0)
      \qbezier(10,0)(15,-10)(20,0)
      \qbezier(0,0)(10,20)(20,0)
    \end{picture}
    \ \
    \simeq
    \ \
    \begin{picture}(20,20)(-10,0)
      \put(0,0){\circle*{4}}
      \put(0,10){\circle*{4}}
      \put(-10,-10){\circle*{4}}
      \put(10,-10){\circle*{4}}
      \qbezier(0,0)(0,0)(0,10)
      \qbezier(0,0)(0,0)(-10,-10)
      \qbezier(0,0)(0,0)(10,-10)
    \end{picture}
  \end{equation}

\item Then for every {\it dessin} it finds all the places where really decomposing rules
  (i.e. [N-1]-rule, [N-2]-rule or [2]-rule) can be applied to it. It generates equations,
  that results of all these decompositions should be the same. These results involve expressions
  for {\it dessins} with fewer edges, which, by assumption, we already know.
  At this point it may happen, that {\it dessin} can not be simplified with [N-1]-rule [N-2]-rule or [2]-rule.
  In this case it is \textit{potentially atomic}.
\item Then, it generates equations that say, that expressions for all {\it dessins} in a cluster
  are related to each other by 1-rule or [N-3]-rule. For every such equation there may be two cases:
  at least one {\it dessin} involved is potentially atomic or there are no potentially atomic {\it dessins}.
  In the latter case we have some equation on {\it dessins} from lower layers.
  In the former case we express one of potentially atomic {\it dessins} using this equation and continue.
\end{itemize}

There are couple of technical details that are important for realization of the algorithm:
\begin{itemize}
\item At every level we need a persistent naming (numbering) of {\it dessins}, because we want to have a way
  to save expressions for them in terms of atomic {\it dessins} and then fetch these expressions.
\item Since some relations may at first look non-trivial, but then completely simplify to zero due
  to identities for $q$-numbers, the CAS is needed to do this polynomial algebra.
  We use Wolfram Mathematica. Partly because it's syntax and idioms are so close to Common Lisp, in which the rest of our program is written.
\end{itemize}

This program is part of our CL-VKNOTS program and can be found on KnoteBook
\cite{knotebook}; as is customary for all the source code nowadays,
it can also be found on GitHub \cite{cl-vknots}.

\section{The results}\label{sec:results}

The more extensive statistics of the program's work is in the appendix \ref{sec:programs-work}.
Here we present the most striking and important features of it.

Up to and including 7 crossings we are not finding any nontrivial relations between atomic {\it dessins}.
Number of atomic {\it dessins} with 0, 1, 2, 3, 4, 5, 6 and 7 edges is equal, respectively,
to 0, 1, 1, 5, 15, 91, 784 and 9160.

Here are first few of them, for small number of edges.

\bigskip

\begin{tabular}{|l|l|}
  \hline
  \# edges & dessins \\ \hline
  1 &
  $\begin{picture}(10,10)(-5,-3)
    \put(0,0){\circle{10}}
    \put(-5,0){\circle*{4}}
  \end{picture}$
  \\ \hline
  2 &
  $\begin{picture}(20,10)(-15,-3)
    \put(0,0){\circle{10}}
    \put(-10,0){\circle{10}}
    \put(-5,0){\circle*{4}}
  \end{picture}$
  \\ \hline
  3 &
  $\begin{picture}(10,22)(0,-9)
    \put(0,0){\circle*{4}}
    \put(0,6){\circle{10}}
    \put(0,-6){\circle{10}}
    \put(5,0){\circle{10}}
  \end{picture}$
   \ \
  $\begin{picture}(20,10)(0,-9)
    \put(0,0){\circle*{4}}
    \put(20,0){\circle*{4}}
    \qbezier(0,0)(10,0)(20,0)
    \qbezier(0,0)(10,10)(20,0)
    \qbezier(0,0)(10,-10)(20,0)
  \end{picture}$
     \ \
  $\begin{picture}(30,10)(-15,-9)
    \put(0,0){\circle*{5}}
    \put(6,0){\circle{10}}
    \put(-6,0){\circle{10}}
    \put(10,0){\circle{20}}
  \end{picture}$
      \ \
  $\begin{picture}(30,10)(-15,-9)
    \put(0,0){\circle*{4}}
    \put(10,0){\circle*{4}}
    \qbezier(0,0)(5,0)(10,0)
    \put(-5,0){\circle{10}}
    \put(15,0){\circle{10}}
  \end{picture}$
        \ \
  $\begin{picture}(30,10)(-15,-9)
    \put(0,0){\circle*{6}}
    \put(-7,0){\circle{10}}
    \put(5,5){\circle{10}}
    \put(3,-6){\circle{10}}
    % \put(15,0){\circle{10}}
  \end{picture}$
  \\ \hline
 & \\ \ldots & \\
\end{tabular}

\bigskip

\textbf{All these {\it dessins} supposedly can be used as new independent variables in HOMFLY polynomial.}
Furthermore, thanks to MOY-formula \cite{MOY} we know, that for planar {\it dessins}
(i.e., non-virtual knots) these additional variables never occur. Therefore,
if HOMFLY polynomial for some planar diagram (involving virtual crossings)
contains these parameters, then this diagram corresponds to \textit{essentially} virtual knot
-- virtual crossings can not be eliminated from this diagram by Reidemeister moves.
Unfortunately, if HOMFLY does not contain these additional parameters, it
does not mean the knot is non-virtual (as appendix \ref{sec:kishino-relatives} shows).

\section{Imposing factorization of composites}\label{sec:factorization-of-composites}

As an example in Appendix \ref{sec:kishino-relatives} illustrates,
inclusion of {\it all} our new parameters breaks factorization property of fundamental HOMFLY
for composite knots.
This property does not follow from topological invariance, still it is common for
all knot polynomials, studied so far -- so one can wish to preserve it.

For this purpose it is sufficient
to impose additional relation on \textit{dessins} -- $1/[N]$-decomposition rule of \cite{MMP12}.

\be
\begin{picture}(300,30)(60,-15)
\put(0,-2){$[N] \cdot D \Bigg ($}
\put(60,0) {
  \put(0,0){\circle{30}}\put(-4,-2){\mbox{$\gamma_1$}}
  \put(80,0){\circle{30}}\put(76,-2){\mbox{$\gamma_2$}}
  \thicklines \linethickness{0.2mm}
  \put(40,0){\circle*{8}}
  %\put(45,5){\vector(0,1){0}}
  \thinlines \linethickness{0.125mm}
  \qbezier(36,0)(35,0)(16,0)
  \qbezier(37,3)(20,20)(13,10)
  \qbezier(37,-3)(20,-20)(13,-10)
  \qbezier(42,0)(45,0)(64,0)
  \qbezier(43,3)(60,20)(67,10)
  \qbezier(43,-3)(60,-20)(67,-10)
}
\put(160,-2){$\Bigg )\ =\ D\Bigg ($}
\put(225,0) {
  \put(0,0){\circle{30}}\put(-4,-2){\mbox{$\gamma_1$}}
  % \put(80,0){\circle{30}}\put(66,-2){\mbox{$\gamma_2$}}
  \thicklines \linethickness{0.2mm}
  \put(40,0){\circle*{8}}
  %\put(45,5){\vector(0,1){0}}
  \thinlines \linethickness{0.125mm}
  \qbezier(36,0)(35,0)(16,0)
  \qbezier(37,3)(20,20)(13,10)
  \qbezier(37,-3)(20,-20)(13,-10)
  %% \qbezier(45,0)(45,0)(65,0)
  %% \qbezier(43,3)(60,20)(67,10)
  %% \qbezier(43,-3)(60,-20)(67,-10)
}
\put(275,-2) {$\Bigg )\ \cdot\ D\Bigg ($}
\put(280,0) {
  % \put(0,0){\circle{30}}\put(-4,-2){\mbox{$\gamma_1$}}
  \put(80,0){\circle{30}}\put(76,-2){\mbox{$\gamma_2$}}
  \thicklines \linethickness{0.2mm}
  \put(40,0){\circle*{8}}
  %\put(45,5){\vector(0,1){0}}
  \thinlines \linethickness{0.125mm}
  %% \qbezier(35,0)(35,0)(15,0)
  %% \qbezier(37,3)(20,20)(13,10)
  %% \qbezier(37,-3)(20,-20)(13,-10)
  \qbezier(43,0)(45,0)(64,0)
  \qbezier(43,3)(60,20)(67,10)
  \qbezier(43,-3)(60,-20)(67,-10)
}
\put(380,-2){$\Bigg )$}
\end{picture}
\label{fact-reduced}
\ee
In other words, if a {\it dessin} has a vertex, which we can cut so that it becomes disconnected,
then we can express dimension of the whole through dimensions of parts.

This considerably reduces the number of {\it dessins}, that are independent -- to

\bigskip

\begin{tabular}{|l|l|}
  \hline
  \# edges & dessins \\ \hline
  1 &
  $\begin{picture}(10,10)(-5,-3)
    \put(0,0){\circle{10}}
    \put(-5,0){\circle*{4}}
  \end{picture}$
  \\ \hline
  2 &
  \\ \hline
  3 &
  $\begin{picture}(10,22)(0,-9)
    \put(0,0){\circle*{4}}
    \put(0,6){\circle{10}}
    \put(0,-6){\circle{10}}
    \put(5,0){\circle{10}}
  \end{picture}$
   \ \
  $\begin{picture}(20,10)(0,-9)
    \put(0,0){\circle*{4}}
    \put(20,0){\circle*{4}}
    \qbezier(0,0)(10,0)(20,0)
    \qbezier(0,0)(10,10)(20,0)
    \qbezier(0,0)(10,-10)(20,0)
  \end{picture}$
   \\ \hline
   4 &
   $\begin{picture}(20,20)(-15,-8)
    \put(1,0){\circle{10}}
    \put(-11,0){\circle{10}}
    \put(-5,6){\circle{10}}
    \put(-5,-6){\circle{10}}
    \put(-5,0){\circle*{6}}
  \end{picture}$
   \ \
  $\begin{picture}(20,10)(0,-9)
    \put(0,0){\circle*{4}}
    \put(20,0){\circle*{4}}
    \qbezier(0,0)(10,0)(20,0)
    \qbezier(0,0)(10,10)(20,0)
    \qbezier(0,0)(10,-20)(20,0)
    \put(-1,-4.5){\circle{10}}
  \end{picture}$
   \ \
  $\begin{picture}(20,32)(-10,-14)
    \put(0,0){\circle*{4}}
    \put(0,6){\circle{10}}
    \put(0,8){\circle{16}}
    \put(0,-6){\circle{10}}
    \put(5,0){\circle{10}}
  \end{picture}$
   \ \
  $\begin{picture}(20,10)(0,-9)
    \put(0,0){\circle*{4}}
    \put(20,0){\circle*{4}}
    \qbezier(0,0)(10,-10)(20,0)
    \qbezier(0,0)(10,10)(20,0)
    \qbezier(0,0)(10,20)(20,0)
    \qbezier(0,0)(10,-20)(20,0)
  \end{picture}$
   \ \
  $\begin{picture}(30,10)(-10,-9)
    \put(0,0){\circle*{4}}
    \put(20,0){\circle*{4}}
    \qbezier(0,0)(10,-10)(20,0)
    \qbezier(0,0)(10,10)(20,0)
    \put(0,-4.5){\circle{10}}
    \put(20,-4.5){\circle{10}}
  \end{picture}$
   \\ \hline
   & \\ \ldots & \\

\end{tabular}

\bigskip

\noindent
but does not eliminate them completely.
Moreover this
number still grows with the number of edges, so the number of new parameters
still seems to be infinite.

\section{Conclusion}\label{sec:conclusion}

In this paper we continued our study of hypercube formalism, applied to virtual knots and links.
We tried to drop as many assumptions on the form of the answer as possible, keeping
only those, that are dictated by skein relation and topological invariance.
In particular, we dropped flip-rule and $1/[N]$-decomposition rule of \cite{MMP12}.

This way, we obtained surprising empirical results. Namely, dimensions
of many hypercube vertices are not fixed by topological invariance
and can, therefore, \textit{be treated as new variables} in HOMFLY polynomial.
Inclusion of these parameters breaks some properties, which are
well-known for HOMFLY polynomials for non-virtual knots but are
subject of discussion for virtual ones, such as factorization of
HOMFLY polynomial for composite knots.
Moreover, even imposing factorization of composites as additional {\it requirement}
(despite it does not follow from topological invariance),
we only {\it diminish} the number of new parameters, while many still seem to survive.

These parameters never appear in HOMFLY polynomials for non-virtual knots
and therefore provide sufficient tool to distinguish essentially virtual knots.
Also, no properties of HOMFLY polynomials for non-virtual knots are broken
as polynomials remain the same.

For $N=2$, i.e. for Jones polynomials, a more powerful approach to knot polynomials is possible.
It uses different skein relation \cite{Kvirt} that, rather than changing the crossing for
inverse one, completely replaces crossing by its two different Morse resolutions.
When applied to virtual knots (what is actually done in \cite{Kvirt}) this approach
has no ambiguities, since after all non-virtual crossings have been eliminated using
this Jones skein relation, all diagrams always completely unknot.
This contrast between HOMFLY-skein and Jones-skein relations is not surprising:
they are already very different for non-virtual knots.
While the latter one allows for recursive calculation of its polynomial,
the former one needs an additional search for unknotting crossings,
which makes calculation an enumerative method, not a recursion.

\bigskip

Since this paper mixes presentation of general theory with empirical results,
i.e. facts with assumptions, the logic is not fully consistent --
what is, however, natural for the subject which is not fully understood.
As we do not have general proofs, we resort to conjectures and checks.
Checks are necessarily done with the help of computer, as the problem
quickly explodes combinatorially. We check our conjectures for
planar diagrams with up to and including 7 non-virtual crossings
and arbitrary number of virtual ones.

We formulate two conjectures, minimal and maximal, which can be briefly
stated as follows:

\begin{conj}
At least one new parameter, consistent with topological invariance, can be put
  into HOMFLY polynomial for virtual knots. It is the value of $q$-dimension of
  the {\it dessin} with one vertex and one   edge:
  $\begin{picture}(20,10)(-5,-2.5)
    \put(0,0){\circle*{5}}
    \put(5,0){\circle{10}}
  \end{picture}$
  This parameter can not be fixed, even if factorization of composites is requested.
\end{conj}
\begin{conj}
An infinite number of free parameters, which are values of $q$-dimensions for
  {\it dessins}, that can not be decomposed using [N-1]-, [N-2]-, [2]-, 1- and [N-3]-rules
  can be put into HOMFLY polynomial for virtual knots, again preserving its topological invariance.
  Many of them are consistent with (i.e. not fixed by) factorization of composites.
\end{conj}

The difference is that empirical evidence in support of the former conjecture looks
overwhelming, while the checks of the latter one are a little less exhaustive.

Extension of these conjectures to {\it colored} HOMFLY of virtual knots and
clarifying of their relation to Chern-Simons/RT approach to HOMFLY polynomials
are important subjects of future research.

\section*{Acknowledgements}

Our work is partly supported by RFBR grants 16-02-01021 (A.M.), 14-02-00627 (And.M. \& A.P.), 15-31-20832-Mol-a-ved (A.M. \& A.P.), by the joint grants 15-52-50041-YaF (A.M. \& A.P.), 15-51-52031-NSC-a (A.M. \& And.M.), 16-51-53034-GFEN (A.M. \& And.M.). Also we are partly supported by the Quantum Topology Lab of Chelyabinsk State University (Russian Federation government grant 14.Z50.31.0020) (And.M.).

\appendix

\section{Kishino knot and its relatives \label{sec:kishino-relatives}}

Kishino knot (4.55 \& 4.56 in the table \cite{virtknottable})
is a composite of two virtual unknots,
which, however, is {\it not} topologically trivial.

\begin{picture}(300,60)(-50,-30)
\qbezier(-10,-5)(30,20)(70,-5)
\qbezier(-10,5)(30,-20)(70,5)
\put(-1,0){\circle{7}}
\put(61,0){\circle{7}}
\qbezier(-10,5)(-30,20)(-10,20)\qbezier(-10,20)(13,20)(16,9)
\qbezier(-10,-5)(-30,-20)(-10,-20)\qbezier(-10,-20)(13,-20)(16,-9)
\qbezier(70,5)(90,20)(70,20)\qbezier(70,20)(47,20)(44,9)
\qbezier(70,-5)(90,-20)(70,-20)\qbezier(70,-20)(47,-20)(44,-9)
\qbezier(17,4)(18,0)(17,-4)
\qbezier(43,4)(42,0)(43,-4)
\put(127,-2){\mbox{$=$}}
\qbezier(190,-5)(230,20)(270,-5)
\qbezier(190,5)(230,-20)(270,5)
\put(199,0){\circle{7}}
\put(261,0){\circle{7}}
\qbezier(190,5)(170,20)(190,20)\qbezier(190,20)(213,20)(216,9)
\qbezier(190,-5)(170,-20)(190,-20)\qbezier(190,-20)(213,-20)(216,-9)
\qbezier(270,5)(290,20)(270,20)\qbezier(270,20)(247,20)(244,9)
\qbezier(270,-5)(290,-20)(270,-20)\qbezier(270,-20)(247,-20)(244,-9)
\qbezier(216,9)(218,0)(216,-9)
\qbezier(244,9)(242,0)(244,-9)
\put(229,7.7){\vector(1,0){2}}
\put(231,-7.7){\vector(-1,0){2}}
\put(190,20){\vector(1,0){2}}
\put(192,-20){\vector(-1,0){2}}
\put(270,20){\vector(1,0){2}}
\put(272,-20){\vector(-1,0){2}}
\put(216.5,6.5){\circle*{4}}
\put(216.5,-6.5){\circle{4}}
\put(243.5,6.5){\circle{4}}
\put(243.5,-6.5){\circle*{4}}
\end{picture}

First we calculate HOMFLY polynomials
for it, and related knots, using well-developed methods of \cite{DM3}.
Then, we use method of this paper to see, how it changes the situation.

\subsection{Gradual quantization (matrix-model) method}
Primary and full classical hypercubes for the main diagram
%with all vertices black

\begin{picture}(300,60)(50,-30)
\qbezier(190,-5)(230,20)(270,-5)
\qbezier(190,5)(230,-20)(270,5)
\put(199,0){\circle{7}}
\put(261,0){\circle{7}}
\qbezier(190,5)(170,20)(190,20)\qbezier(190,20)(213,20)(216,9)
\qbezier(190,-5)(170,-20)(190,-20)\qbezier(190,-20)(213,-20)(216,-9)
\qbezier(270,5)(290,20)(270,20)\qbezier(270,20)(247,20)(244,9)
\qbezier(270,-5)(290,-20)(270,-20)\qbezier(270,-20)(247,-20)(244,-9)
\qbezier(216,9)(218,0)(216,-9)
\qbezier(244,9)(242,0)(244,-9)
\put(229,7.7){\vector(1,0){2}}
\put(231,-7.7){\vector(-1,0){2}}
\put(190,20){\vector(1,0){2}}
\put(192,-20){\vector(-1,0){2}}
\put(270,20){\vector(1,0){2}}
\put(272,-20){\vector(-1,0){2}}
%\put(216.5,6.5){\circle*{4}}
%\put(216.5,-6.5){\circle*{4}}
%\put(243.5,6.5){\circle*{4}}
%\put(243.5,-6.5){\circle*{4}}
\end{picture}

\noindent
calculated by the rules of \cite{virtHOMFLY},
are

\bigskip

\centerline{
$
\begin{array}{ccccccccc}
&&&&1       \\
&&2&&3&&2 \\
&&2&&3&&2 \\
\boxed{1}&&&&&&&&1 \\
&&2&&3&&2 \\
&&2&&3&&2 \\
&&&&1
\end{array}
\ \ \ \ \ \ \longrightarrow \ \ \ \
\begin{array}{ccccccccc}
&&&&2N-2N^2       \\
&&N-N^2&&N^3-2N^2+N&&2N^3-4N^2+2N \\
&&N-N^2&&N^3-2N^2+N&&2N^3-4N^2+2N \\
\boxed{N}&&&&&&&&4N^3-8N^2+4N \\
&&N-N^2&&N^3-2N^2+N&&2N^3-4N^2+2N \\
&&N-N^2&&N^3-2N^2+N&&2N^3-4N^2+2N \\
&&&&2N-2N^2
\end{array}
$
}

\bigskip

Quantization in this case is straightforward and the full hypercube is:
\be
\begin{array}{ccccccccc}
&&&&\boxed{\boxed{-[2][N][N-1]}}       \\
&&-[N][N-1]&&[N][N-1]^2 &&[2][N][N-1]^2 \\
&&-[N][N-1]&&[N][N-1]^2&&[2][N][N-1]^2 \\
\boxed{[N]}&&&&&&&&[2]^2[N][N-1]^2 \\
&&-[N][N-1]&&[N][N-1]^2&&[2][N][N-1]^2 \\
&&-[N][N-1]&&\boxed{\boxed{\boxed{[N][N-1]^2}}}&&[2][N][N-1]^2 \\
&&&&-[2][N][N-1]
\end{array}
\nn
\ee

Starting from the main (Seifert) vertex, marked by a single box, we obtain for

\begin{equation} \label{eq:seifert-kishino-pic}
\begin{picture}(300,60)(100,-20)
\qbezier(190,-5)(230,20)(270,-5)
\qbezier(190,5)(230,-20)(270,5)
\put(199,0){\circle{7}}
\put(261,0){\circle{7}}
\qbezier(190,5)(170,20)(190,20)\qbezier(190,20)(213,20)(216,9)
\qbezier(190,-5)(170,-20)(190,-20)\qbezier(190,-20)(213,-20)(216,-9)
\qbezier(270,5)(290,20)(270,20)\qbezier(270,20)(247,20)(244,9)
\qbezier(270,-5)(290,-20)(270,-20)\qbezier(270,-20)(247,-20)(244,-9)
\qbezier(216,9)(218,0)(216,-9)
\qbezier(244,9)(242,0)(244,-9)
\put(229,7.7){\vector(1,0){2}}
\put(231,-7.7){\vector(-1,0){2}}
\put(190,20){\vector(1,0){2}}
\put(192,-20){\vector(-1,0){2}}
\put(270,20){\vector(1,0){2}}
\put(272,-20){\vector(-1,0){2}}
\put(216.5,6.5){\circle*{4}}
\put(216.5,-6.5){\circle*{4}}
\put(243.5,6.5){\circle*{4}}
\put(243.5,-6.5){\circle*{4}}
\end{picture}
\end{equation}

\be
q^{4N-4}[N]\left\{1+ 4q[N-1]+4q^2[N-1]^2-2q^2[2][N-1]-4q^3[2][N-1]^2+q^4[2]^2[N-1]^2
\right\}
= \nn \\
= q^{4(N-1)}[N]\Big(1-q^2(q-q^{-1})[N-1]\Big)^2
= [N]\Big(q^{N+1}+q^{2N-2}-q^{3N-1}\Big)^2
=[N]\cdot\Big( v^{[2.1]}_{_\Box}\Big)^2
\ee
as one should expect for a composite of the two virtual knots $2.1$.

Similarly for

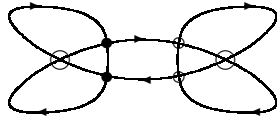
\begin{picture}(300,60)(50,-30)
\qbezier(190,-5)(230,20)(270,-5)
\qbezier(190,5)(230,-20)(270,5)
\put(199,0){\circle{7}}
\put(261,0){\circle{7}}
\qbezier(190,5)(170,20)(190,20)\qbezier(190,20)(213,20)(216,9)
\qbezier(190,-5)(170,-20)(190,-20)\qbezier(190,-20)(213,-20)(216,-9)
\qbezier(270,5)(290,20)(270,20)\qbezier(270,20)(247,20)(244,9)
\qbezier(270,-5)(290,-20)(270,-20)\qbezier(270,-20)(247,-20)(244,-9)
\qbezier(216,9)(218,0)(216,-9)
\qbezier(244,9)(242,0)(244,-9)
\put(229,7.7){\vector(1,0){2}}
\put(231,-7.7){\vector(-1,0){2}}
\put(190,20){\vector(1,0){2}}
\put(192,-20){\vector(-1,0){2}}
\put(270,20){\vector(1,0){2}}
\put(272,-20){\vector(-1,0){2}}
\put(216.5,6.5){\circle*{4}}
\put(216.5,-6.5){\circle*{4}}
\put(243.5,6.5){\circle{4}}
\put(243.5,-6.5){\circle{4}}
\end{picture}

\noindent
when initial is the double-boxed vertex,
\be
q^{-2}[N]\left\{-[2][N-1]-q\Big(2[2][N-1]^2-2[N-1]\Big)+
q^2\Big(1+[2]^2[N-1]^2+4[N-1]^2\Big)-
\right.\nn \\ \left.  \!\!\!\!\!\!  -
q^3\Big(2[2][N-1]^2-2[N-1]\Big)-q^4[2][N-1]
\right\}  \
=\  [N]\Big(1-[2][N-1](q-q^{-1})^2+[N-1]^2(q-q^{-1})^2\Big) = \nn \\
 = [N]\Big(1-q^2(q-q^{-1})[N-1]\Big)\Big(1-q^{-2}(q-q^{-1})[N-1]\Big)
= [N]\cdot v^{[2.1]}_{_\Box}(q)\cdot v^{[2.1]}_{_\Box}(q^{-1})
\ee
-- what is again the expected answer for a composite of $2.1$ and its
copy with inverted orientation.

Thus it comes with no surprise that the Kishino knot itself, which
is a composite of two virtual unknots, has a trivial HOMFLY.
Indeed, taking the triple-boxed vertex as initial, we get for either

\begin{equation} \label{eq:kishino-pic-1}
\begin{picture}(300,65)(0,-30)
  \qbezier(-10,-5)(30,20)(70,-5)
  \qbezier(-10,5)(30,-20)(70,5)
  \put(-1,0){\circle{7}}
  \put(61,0){\circle{7}}
  \qbezier(-10,5)(-30,20)(-10,20)\qbezier(-10,20)(13,20)(16,9)
  \qbezier(-10,-5)(-30,-20)(-10,-20)\qbezier(-10,-20)(13,-20)(16,-9)
  \qbezier(70,5)(90,20)(70,20)\qbezier(70,20)(47,20)(44,9)
  \qbezier(70,-5)(90,-20)(70,-20)\qbezier(70,-20)(47,-20)(44,-9)
  \qbezier(17,4)(18,0)(17,-4)
  \qbezier(43,4)(42,0)(43,-4)
  \put(127,-2){\mbox{$=$}}
  \qbezier(190,-5)(230,20)(270,-5)
  \qbezier(190,5)(230,-20)(270,5)
  \put(199,0){\circle{7}}
  \put(261,0){\circle{7}}
  \qbezier(190,5)(170,20)(190,20)\qbezier(190,20)(213,20)(216,9)
  \qbezier(190,-5)(170,-20)(190,-20)\qbezier(190,-20)(213,-20)(216,-9)
  \qbezier(270,5)(290,20)(270,20)\qbezier(270,20)(247,20)(244,9)
  \qbezier(270,-5)(290,-20)(270,-20)\qbezier(270,-20)(247,-20)(244,-9)
  \qbezier(216,9)(218,0)(216,-9)
  \qbezier(244,9)(242,0)(244,-9)
  \put(229,7.7){\vector(1,0){2}}
  \put(231,-7.7){\vector(-1,0){2}}
  \put(190,20){\vector(1,0){2}}
  \put(192,-20){\vector(-1,0){2}}
  \put(270,20){\vector(1,0){2}}
  \put(272,-20){\vector(-1,0){2}}
  \put(216.5,6.5){\circle*{4}}
  \put(216.5,-6.5){\circle{4}}
  \put(243.5,6.5){\circle{4}}
  \put(243.5,-6.5){\circle*{4}}
\end{picture}
\end{equation}

\noindent
or

\begin{equation} \label{eq:kishino-pic-2}
\begin{picture}(300,60)(0,-30)
\qbezier(-10,-5)(20,13)(41,7)\qbezier(47,6)(60,3)(70,-5)
\qbezier(-10,5)(20,-13)(41,-7)\qbezier(47,-6)(60,-3)(70,5)
\put(-1,0){\circle{7}}
\put(61,0){\circle{7}}
\qbezier(-10,5)(-30,20)(-10,20)\qbezier(-10,20)(13,20)(16,9)
\qbezier(-10,-5)(-30,-20)(-10,-20)\qbezier(-10,-20)(13,-20)(16,-9)
\qbezier(70,5)(90,20)(70,20)\qbezier(70,20)(47,20)(44,9)
\qbezier(70,-5)(90,-20)(70,-20)\qbezier(70,-20)(47,-20)(44,-9)
\qbezier(17,4)(18,0)(17,-4)
\qbezier(44,9)(42,0)(44,-9)
\put(127,-2){\mbox{$=$}}
\qbezier(190,-5)(230,20)(270,-5)
\qbezier(190,5)(230,-20)(270,5)
\put(199,0){\circle{7}}
\put(261,0){\circle{7}}
\qbezier(190,5)(170,20)(190,20)\qbezier(190,20)(213,20)(216,9)
\qbezier(190,-5)(170,-20)(190,-20)\qbezier(190,-20)(213,-20)(216,-9)
\qbezier(270,5)(290,20)(270,20)\qbezier(270,20)(247,20)(244,9)
\qbezier(270,-5)(290,-20)(270,-20)\qbezier(270,-20)(247,-20)(244,-9)
\qbezier(216,9)(218,0)(216,-9)
\qbezier(244,9)(242,0)(244,-9)
\put(229,7.7){\vector(1,0){2}}
\put(231,-7.7){\vector(-1,0){2}}
\put(190,20){\vector(1,0){2}}
\put(192,-20){\vector(-1,0){2}}
\put(270,20){\vector(1,0){2}}
\put(272,-20){\vector(-1,0){2}}
\put(216.5,6.5){\circle*{4}}
\put(216.5,-6.5){\circle{4}}
\put(243.5,6.5){\circle*{4}}
\put(243.5,-6.5){\circle{4}}
\end{picture}
\end{equation}

\be
& q^{-2}[N] \left\{ [N-1]^2 - q \lb - 2 [N-1] + 2 [2][N-1]^2\rb \right. & \\ \nn
& \left. + q^2 \lb 1 + 2 [2] (-) [N-1] + 2 [N-1]^2 + [2]^2 [N-1]^2 \rb
- q^3 \lb 2 (-) [N-1] + 2 [2] [N-1]^2 \rb
+ q^4 [N-1]^2
\right\}
& \\ \nn & = [N]
\ee

\subsection{Axiomatic (MOY-relations) method}

Let's now calculate HOMFLY for Kishino knot and related knots using techniques discussed
in this paper. We start with the knot in the Seifert vertex
(picture \eqref{eq:seifert-kishino-pic}). Corresponding {\it dessin d'enfant} is

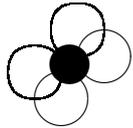
\begin{equation} \nn
  \begin{picture}(40,40)(0,-20)
    \put(0,0){\circle*{20}}
    \put(-3,-3){
      \put(0,-10){\circle{20}}
      % \thicklines
      \qbezier(0,0)(0,-10)(-10,-10) \qbezier(-10,-10)(-20,-10)(-20,0)
      \qbezier(-20,0)(-20,10)(-10,10) \qbezier(-10,10)(0,10)(0,0)
    }
    \put(3,3){
      \put(10,0){\circle{20}}
      % \thicklines
      \put(10,10){
        \qbezier(0,0)(0,-10)(-10,-10) \qbezier(-10,-10)(-20,-10)(-20,0)
        \qbezier(-20,0)(-20,10)(-10,10) \qbezier(-10,10)(0,10)(0,0)
      }
    }
  \end{picture}
\end{equation}

\noindent hence HOMFLY polynomial, according to the amputation formula \eqref{eq:hypercube-formula}, is

\be
  H_{\text{Seifert}} = & q^{4 N - 4}
  \left (
  q^4 \ \
  \begin{picture}(40,40)(-20,0)
    \put(0,0){\circle*{20}}
    \put(-3,-3){
      \put(0,-10){\circle{20}}
      \qbezier(0,0)(0,-10)(-10,-10) \qbezier(-10,-10)(-20,-10)(-20,0)
      \qbezier(-20,0)(-20,10)(-10,10) \qbezier(-10,10)(0,10)(0,0)
    }
    \put(3,3){
      \put(10,0){\circle{20}}
      \put(10,10){
        \qbezier(0,0)(0,-10)(-10,-10) \qbezier(-10,-10)(-20,-10)(-20,0)
        \qbezier(-20,0)(-20,10)(-10,10) \qbezier(-10,10)(0,10)(0,0)
      }
    }
  \end{picture}
  \ \ - 4 q^3 \ \
  \begin{picture}(30,40)(-15,0)
    \put(0,0){\circle*{20}}
    \put(-3,-3){
      \put(0,-10){\circle{20}}
    }
    \put(3,3){
      \put(10,0){\circle{20}}
      \put(10,10){
        \qbezier(0,0)(0,-10)(-10,-10) \qbezier(-10,-10)(-20,-10)(-20,0)
        \qbezier(-20,0)(-20,10)(-10,10) \qbezier(-10,10)(0,10)(0,0)
      }
    }
  \end{picture}
  \ \ + 4 q^2
  \begin{picture}(30,40)(-15,0)
    \put(0,0){\circle*{20}}
    \put(-3,-3){
      \put(0,-10){\circle{20}}
    }
    \put(3,3){
      \put(10,0){\circle{20}}
    }
  \end{picture}
  \ \ + 2 q^2
  \begin{picture}(40,40)(-20,0)
    \put(0,0){\circle*{20}}
    \put(3,3){
      \put(10,0){\circle{20}}
      \put(10,10){
        \qbezier(0,0)(0,-10)(-10,-10) \qbezier(-10,-10)(-20,-10)(-20,0)
        \qbezier(-20,0)(-20,10)(-10,10) \qbezier(-10,10)(0,10)(0,0)
      }
    }
  \end{picture}
  \ \ - 4 q
  \begin{picture}(30,40)(-15,0)
    \put(0,0){\circle*{20}}
    \put(3,3){
      \put(10,0){\circle{20}}
    }
  \end{picture}
  \ \ +
  \begin{picture}(20,40)(-10,0)
    \put(0,0){\circle*{20}}
  \end{picture}
  \right ) & \\ \nn
  = & q^{4 N - 4} \lb [N] + 2 q^2
  \lb q - \frac{1}{q} \rb
  \begin{picture}(10,10)(-5,-2.5)
    \put(0,0){\circle*{5}}
    \put(5,0){\circle{10}}
  \end{picture}
  \ \ + q^4 \lb q - \frac{1}{q} \rb^2
  \begin{picture}(20,10)(-10,-2.5)
    \put(0,0){\circle*{5}}
    \put(6,0){\circle{10}}
    \put(-6,0){\circle{10}}
  \end{picture}
 \rb
\ee

\noindent
On the other hand,
the planar diagram of a virtual trefoil (2.1 in the table \cite{virtknottable}),
has {\it dessin d'enfant}

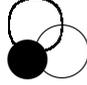
\begin{equation} \nn
  \begin{picture}(40,40)(0,-20)
    \put(0,0){\circle*{20}}
    \put(3,3){
      \put(10,0){\circle{20}}
      \put(10,10){
        \qbezier(0,0)(0,-10)(-10,-10) \qbezier(-10,-10)(-20,-10)(-20,0)
        \qbezier(-20,0)(-20,10)(-10,10) \qbezier(-10,10)(0,10)(0,0)
      }
    }
  \end{picture}
\end{equation}

\noindent
hence for its HOMFLY polynomial we get

\be
H_{\text{2.1}} = q^{2 N - 2} \lb [N] + q^2 \lb q - \frac{1}{q} \rb
\begin{picture}(15,10)(-5,-2.5)
    \put(0,0){\circle*{5}}
    \put(5,0){\circle{10}}
  \end{picture}
\rb
\ee

\noindent so the factorization property for composites
$H_{\text{Seifert}} = 1/[N] H_{\text{2.1}}^2$ is no longer true, unless, of course,
one imposes \textit{by hand} (we have not found any evidence, that it's required by the topological invariance) that
\be
  \begin{picture}(20,10)(-10,-2.5)
    \put(0,0){\circle*{5}}
    \put(6,0){\circle{10}}
    \put(-6,0){\circle{10}}
  \end{picture} = \frac{1}{[N]}
  \begin{picture}(15,10)(-5,-2.5)
    \put(0,0){\circle*{5}}
    \put(5,0){\circle{10}}
  \end{picture}^2
\ee

The two versions of Kishino knot
(pictures \eqref{eq:kishino-pic-1} and \eqref{eq:kishino-pic-2})
have the following {\it dessins d'enfant}
(dotted edges correspond to white vertices of planar diagram)

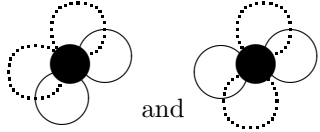
\begin{equation} \nn
  \begin{picture}(40,40)(-20,-20)
    \put(0,0){\circle*{20}}
    \put(-3,-3){
      \put(0,-10){\circle{20}}
      \thicklines
      \qbezier[7](0,0)(0,-10)(-10,-10) \qbezier[7](-10,-10)(-20,-10)(-20,0)
      \qbezier[7](-20,0)(-20,10)(-10,10) \qbezier[7](-10,10)(0,10)(0,0)
    }
    \put(3,3){
      \put(10,0){\circle{20}}
      \thicklines
      \put(10,10){
        \qbezier[7](0,0)(0,-10)(-10,-10) \qbezier[7](-10,-10)(-20,-10)(-20,0)
        \qbezier[7](-20,0)(-20,10)(-10,10) \qbezier[7](-10,10)(0,10)(0,0)
      }
    }
  \end{picture}
  \ \ \text{and} \ \
  \begin{picture}(40,40)(-20,-20)
    \put(0,0){\circle*{20}}
    \put(-3,-3){\put(-10,0){\circle{20}}}
    \put(-12,-14) {
      \thicklines
      \qbezier[7](0,0)(0,10)(10,10) \qbezier[7](10,10)(20,10)(20,0)
      \qbezier[7](20,0)(20,-10)(10,-10) \qbezier[7](10,-10)(0,-10)(0,0)
    }
    \put(3,3){
      \put(10,0){\circle{20}}
      \thicklines
      \put(10,10){
        \qbezier[7](0,0)(0,-10)(-10,-10) \qbezier[7](-10,-10)(-20,-10)(-20,0)
        \qbezier[7](-20,0)(-20,10)(-10,10) \qbezier[7](-10,10)(0,10)(0,0)
      }
    }
  \end{picture}
\end{equation}

The combinatorics of $q$-factors in amputation formula \eqref{eq:hypercube-formula}
is the same for both of them and reads
(here we omit difference between solid/dotted lines, since we've already written
out all $q$-weights explicitly)

\be
  H_{\text{Kishino}} = &
  \left (
  \ \
  \begin{picture}(40,40)(-20,0)
    \put(0,0){\circle*{20}}
    \put(-3,-3){
      \put(0,-10){\circle{20}}
      \qbezier(0,0)(0,-10)(-10,-10) \qbezier(-10,-10)(-20,-10)(-20,0)
      \qbezier(-20,0)(-20,10)(-10,10) \qbezier(-10,10)(0,10)(0,0)
    }
    \put(3,3){
      \put(10,0){\circle{20}}
      \put(10,10){
        \qbezier(0,0)(0,-10)(-10,-10) \qbezier(-10,-10)(-20,-10)(-20,0)
        \qbezier(-20,0)(-20,10)(-10,10) \qbezier(-10,10)(0,10)(0,0)
      }
    }
  \end{picture}
  \ \ - 2 \lb q + \frac{1}{q} \rb \ \
  \begin{picture}(20,40)(-5,0)
    \put(0,0){\circle*{20}}
    \put(-3,-3){
      \put(0,-10){\circle{20}}
    }
    \put(3,3){
      \put(10,0){\circle{20}}
      \put(10,10){
        \qbezier(0,0)(0,-10)(-10,-10) \qbezier(-10,-10)(-20,-10)(-20,0)
        \qbezier(-20,0)(-20,10)(-10,10) \qbezier(-10,10)(0,10)(0,0)
      }
    }
  \end{picture}
  \ \ + \lb q^2 + 2 + \frac{1}{q^2} \rb
  \begin{picture}(20,40)(-5,0)
    \put(0,0){\circle*{20}}
    \put(-3,-3){
      \put(0,-10){\circle{20}}
    }
    \put(3,3){
      \put(10,0){\circle{20}}
    }
  \end{picture}
  \ \ + 2
  \begin{picture}(20,40)(-10,0)
    \put(0,0){\circle*{20}}
    \put(3,3){
      \put(10,0){\circle{20}}
      \put(10,10){
        \qbezier(0,0)(0,-10)(-10,-10) \qbezier(-10,-10)(-20,-10)(-20,0)
        \qbezier(-20,0)(-20,10)(-10,10) \qbezier(-10,10)(0,10)(0,0)
      }
    }
  \end{picture}
  \ \ - 2 \lb q + \frac{1}{q} \rb
  \begin{picture}(30,40)(-15,0)
    \put(0,0){\circle*{20}}
    \put(3,3){
      \put(10,0){\circle{20}}
    }
  \end{picture}
  \ \ +
  \begin{picture}(20,40)(-10,0)
    \put(0,0){\circle*{20}}
  \end{picture}
  \right ) \ \ \ & \\ \nn
  = &
  [2]^2
  \begin{picture}(20,10)(-10,-2.5)
    \put(0,0){\circle*{5}}
    \put(6,0){\circle{10}}
    \put(-6,0){\circle{10}}
  \end{picture}
  - 2 [2]^2
  \begin{picture}(20,10)(-10,-2.5)
    \put(0,0){\circle*{5}}
    \put(6,0){\circle{10}}
    \put(-6,0){\circle{10}}
  \end{picture}
  + [2]^2
  \begin{picture}(20,10)(-10,-2.5)
    \put(0,0){\circle*{5}}
    \put(6,0){\circle{10}}
    \put(-6,0){\circle{10}}
  \end{picture}
  + 2 [2]
  \begin{picture}(20,10)(-5,-2.5)
    \put(0,0){\circle*{5}}
    \put(5,0){\circle{10}}
  \end{picture}
  - 2 [2]
  \begin{picture}(20,10)(-5,-2.5)
    \put(0,0){\circle*{5}}
    \put(5,0){\circle{10}}
  \end{picture}
   + [N] = [N]
\ee

\noindent
Thus our new independent parameters cancel from HOMFLY polynomial. This shows
that appearance of these new parameters is sufficient for a knot to be essentially
virtual, but not necessary.

\section{Results of the program's work \label{sec:programs-work}}

%% What the program gives us, can be summarized in the following

%% \begin{conj}
%% There are no non-trivial relations and there are lots of atomic dessins.
%% \end{conj}

Even with computer program we are able to get to {\it dessins} with only up to and including 7 edges.
This is because the number of distinct (connected) {\it dessins} with given number of edges
grows as

\bigskip

\begin{tabular}{|l|c|c|c|c|c|c|c|c|}
  \hline
\# edges & 0 & 1 & 2 & 3 & 4 & 5 & 6 & 7 \\ \hline
\# dessins & 1 & 2 & 5 & 20 & 107 & 870 & 9436 & 122840 \\ \hline
\end{tabular}

\bigskip

\noindent
The number of relations, that come from application of [N-1]- [N-2]- or [2]-rules
to different places of the same {\it dessin} is, respectively

\bigskip

\begin{tabular}{|l|c|c|c|c|c|c|c|c|}
  \hline
\# edges & 0 & 1 & 2 & 3 & 4 & 5 & 6 & 7 \\ \hline
\# simple rels & 0 & 1 & 4 & 14 & 86 & 729 & 8385 & 113399 \\ \hline
\end{tabular}

\bigskip

The number of {\it dessins}, decompositions of which give the given number of equations
\begin{verbatim}
1: 1 -> 1
2:             2 -> 2
3: 1 -> 5,     2 -> 3,     3 -> 1
4: 1 -> 32,    2 -> 16,    3 -> 6,    4 -> 1
5: 1 -> 265,   2 -> 136,   3 -> 49,   4 -> 10,   5 -> 1
6: 1 -> 2793,  2 -> 1588,  3 -> 580,  4 -> 140,  5 -> 22,  6 -> 1
7: 1 -> 36068, 2 -> 20996, 3 -> 8175, 4 -> 2162, 5 -> 385, 6 -> 39, 7 -> 1
\end{verbatim}

However, all these relations actually \textbf{simplify to zero}, thanks to identities between
$q$-numbers (i.e. coefficient in front of every monomial in atomic {\it dessins} is separately equal to zero).

\bigskip

The number of cluster relations (i.e. the ones which relate {\it dessins} by [N-3]- or 1-rule) is

\bigskip

\begin{tabular}{|l|c|c|c|c|c|c|c|}
  \hline
\# edges & 1 & 2 & 3 & 4 & 5 & 6 & 7 \\ \hline
\# cluster rels & 0 & 0 & 2 & 18 & 250 & 3415 & 49316 \\ \hline
\end{tabular}

\bigskip

\noindent
Among them, the following number contains potentially atomic {\it dessins} and hence
is used to express these potentially atomic {\it dessins} through sub-{\it dessins}

\bigskip

\begin{tabular}{|l|c|c|c|c|c|c|c|}
  \hline
\# edges & 1 & 2 & 3 & 4 & 5  & 6   & 7    \\ \hline
\# rels  & 0 & 0 & 0 & 3 & 46 & 630 & 8631 \\ \hline
\end{tabular}

\bigskip

All other cluster relations connect two {\it dessins}, each of which is independently simplifiable
by [N-1]- [N-2] or [2]-rule, and hence in principle could've given some relation between
sub-{\it dessins}, but all of them \textbf{simplify to zero}.

\bigskip

The number of {\it dessins}, that are expressed either through sub-{\it dessins}, or, via cluster relations,
through {\it dessins} on the same level, is

\bigskip

\begin{tabular}{|l|c|c|c|c|c|c|c|c|}
  \hline
\# edges & 0 & 1 & 2 & 3 & 4 & 5 & 6 & 7 \\ \hline
\# molecular dessins & 1 & 1 & 4 & 15 & 92 & 779 & 8652 & 113680 \\ \hline
\end{tabular}

\bigskip

\noindent
Hence, the number of atomic {\it dessins} at each level (i.e. number of potentially atomic minus
the ones which were expressed using cluster relations on this level) is

\bigskip

\begin{tabular}{|l|c|c|c|c|c|c|c|c|} \hline
\# edges & 0 & 1 & 2 & 3 & 4 & 5 & 6 & 7 \\ \hline
\# atomic dessins & 0 & 1 & 1 & 5 & 15 & 91 & 784 & 9160 \\ \hline
\end{tabular}

\bigskip

In result, it looks like we have a lot of atomic, indecomposable {\it dessins}
and their number grows fast with the number of edges.
However, it still may be the case that we are not looking into the structure deep enough.
Since number of different {\it dessins} with a given number of edges also grows fast, it
may be that at certain point (say, 20 edges) they start to provide many constraints
on values of {\it dessins} with lower number of edges.
Say, 20-edge {\it dessins} could constrain completely
2-edge {\it dessins}, 21-edge {\it dessins} -- 3-edge {\it dessins} and so on.
Since there is no hope to address the issue by brute force (due to combinatorial explosion),
a more intelligent method is needed.
This is, however, beyond the scope of the present, largely empirical, investigation.

\section{Hypercube construction in symmetric channel
\label{Appsymm}}

In this section we develop hypercube construction slightly differently.
The difference is that instead of projector onto antisymmetric representation $P_{[1,1]}$
central role is played by projector onto symmetric representation $P_{[2]}$.
If one does not keep in mind, that HOMFLY admit Khovanov-Rozansky categorification,
this view should be no worse than traditional, antisymmetric one.
This allows us to show, that some choices, that seem canonical and rigid
in the antisymmetric picture are not so fixed.

Namely, we can start with formula for ${\cal R}$-matrix through symmetric projector, instead
of symmetric one
\be
{\cal R} = [2] P_{[2]} - q I \otimes I  \\ \nn
{\cal R}^{-1} = [2] P_{[2]} - \frac{1}{q} I \otimes I
\ee

Then, instead of hypercube formula \eqref{eq:hypercube-formula} one gets
the following analogous formula
\be \label{eq:hypercube-formula-app}
\  H^{{\cal L}_c}_{_\Box}
=   \lb q^N (-q) \rb^{\lb n_{\bullet}(\GLC)-n_{\circ}(\GLC) \rb}
\sum_{\gamma\subseteq \GLC} \lb-\frac{1}{q}\rb^{n_{\bullet}(\gamma) - n_\circ(\gamma)}
\cdot D_\gamma(q,N)
\phantom{5^{5^{5^{5^{5^5}}}}}\!\!\!\!\!\!\!\!\!\!\!\!,
\ee
where now edges of fat graphs correspond to $[2] P_{[2]}$.

Invariance with respect to Reidemeister moves, together with
factorization and normalization properties (see section \ref{sec:moy-rels})
lead to the following relations on these, symmetric, fat graphs

\be \label{eq:n-1-rule-sym} \label{eq:moy-rels-sym-b}
&  \begin{picture}(300,50)(0,-10)
\put(100,0){
    \linethickness{0.4mm} \thicklines
    \qbezier(0,0)(20,20)(0,40)
    \put(0,40){\vector(-1,1){0}}
    \put(35,10){\vector(1,0){0}}
    \put(30,20){\circle{20}}
        {\linethickness{0.15mm} \put(10,20){\line(1,0){10}}}
        \put(50,17) {$\simeq\ \ [N+1]$}
        \put(110,0){\vector(0,1){40}}
}
  \end{picture}
\ee

\be \label{eq:2-rule-sym}
& \begin{picture}(300,60)(-80,-20)
    \linethickness{0.4mm} \thicklines
    \qbezier(0,0)(20,20)(0,40)
    \put(0,40){\vector(-1,1){0}}
    \put(40,0){\qbezier(0,0)(-20,20)(0,40)}
    \put(40,40){\vector(1,1){0}}
        {\linethickness{0.15mm}
          \put(8,30){\line(1,0){24}}
          \put(8,10){\line(1,0){24}}
        }
        \put(50,17) {$\simeq\ \ [2]$}
        \put(80,0){
          \linethickness{0.4mm} \thicklines
          \qbezier(0,0)(20,20)(0,40)
          \put(0,40){\vector(-1,1){0}}
          \put(40,0){\qbezier(0,0)(-20,20)(0,40)}
          \put(40,40){\vector(1,1){0}}
              {\linethickness{0.15mm}
                \put(10,20){\line(1,0){20}}
              }
        }
  \end{picture}
\ee

\be \label{eq:n-2-rule-sym}
& \begin{picture}(300,60)(-80,-20)
    \linethickness{0.4mm} \thicklines
    \put(-40,0){\qbezier(0,0)(20,20)(0,40) \put(0,40){\vector(-1,1){0}}}
    \put(40,0){\qbezier(0,0)(-20,20)(0,40)}
    \put(40,0){\vector(1,-1){0}}
    \put(-40,0){\linethickness{0.15mm}
      \put(10,20){\line(1,0){20}}
    }
    \put(0,0){\linethickness{0.15mm}
      \put(10,20){\line(1,0){20}}
    }
    \put(0,20){\circle{20} \put(5,10){\vector(1,0){0}} \put(-5,-10){\vector(-1,0){0}}}

    \put(50,17) {$\simeq\ \ [N+2]$}
    \put(100,0) {
      \linethickness{0.4mm} \thicklines
      \qbezier(0,0)(20,20)(0,40)
      \put(0,40){\vector(-1,1){0}}
      \put(40,0){\qbezier(0,0)(-20,20)(0,40)}
      \put(40,0){\vector(1,-1){0}}
    }
    \put(150,17){$+$}
    \put(165,0) {
      \linethickness{0.4mm} \thicklines
      \qbezier(0,0)(20,20)(40,0)
      \put(0,40){\vector(-1,1){0}}
      \put(40,40){\qbezier(0,0)(-20,-20)(-40,0)}
      \put(40,0){\vector(1,-1){0}}
    }
  \end{picture}
\ee

\be \label{eq:n-3-rule-sym}
& \begin{picture}(300,50)(-10,-10)
    \put(-40,0) {
      \linethickness{0.4mm} \thicklines
      \qbezier(0,0)(0,20)(-20,20) \put(-20,20){\vector(-1,0){0}}
      \put(20,0){\qbezier(0,0)(0,20)(20,20) \put(0,0){\vector(0,-1){0}}}
      \put(20,40){\qbezier(0,0)(-10,-10)(-20,0) \put(0,0){\vector(2,1){0}}}
      \put(10,20){\circle{16}}
      \put(10,0){
        \linethickness{0.15mm} \thinlines
        \qbezier(-12,11)(-12,11)(-6,15)
      }
      \put(10,0){
        \linethickness{0.15mm} \thinlines
        \qbezier(12,11)(12,11)(6,15)
      }
      \put(10,28){
        \linethickness{0.15mm} \thinlines
        \qbezier(0,0)(0,5)(0,7)
      }
    }
    \put(10,17){$-\ [N+3]$}
    \put(80,0) {
      \linethickness{0.4mm} \thicklines
      \qbezier(0,0)(0,20)(-20,20) \put(-20,20){\vector(-1,0){0}}
      \put(20,0){\qbezier(0,0)(0,20)(20,20) \put(0,0){\vector(0,-1){0}}}
      \put(20,40){\qbezier(0,0)(-10,-10)(-20,0) \put(0,0){\vector(2,1){0}}}
    }
    \put(130,17){$\simeq$}
    \put(210,0) {
      \put(-40,0) {
        \linethickness{0.4mm} \thicklines
        \qbezier(0,40)(0,20)(-20,20) \put(-20,20){\vector(-1,0){0}}
        \put(20,0){\qbezier(0,40)(0,20)(20,20) \put(0,40){\vector(0,1){0}}}
        \put(20,0){\qbezier(0,0)(-10,10)(-20,0) \put(0,0){\vector(2,-1){0}}}
        \put(10,20){\circle{16}}
        \put(10,40){
          \linethickness{0.15mm} \thinlines
          \qbezier(-12,-11)(-12,-11)(-6,-15)
        }
        \put(10,40){
          \linethickness{0.15mm} \thinlines
          \qbezier(12,-11)(12,-11)(6,-15)
        }
        \put(10,5){
          \linethickness{0.15mm} \thinlines
          \qbezier(0,0)(0,5)(0,7)
        }
      }
      \put(10,17){$-\ [N+3]$}
      \put(80,0) {
        \linethickness{0.4mm} \thicklines
        \qbezier(0,40)(0,20)(-20,20) \put(-20,20){\vector(-1,0){0}}
        \put(20,0){\qbezier(0,40)(0,20)(20,20) \put(0,40){\vector(0,1){0}}}
        \put(20,0){\qbezier(0,0)(-10,10)(-20,0) \put(0,0){\vector(2,-1){0}}}
      }
    }
  \end{picture}
\ee

\be \label{eq:1-rule-sym}  \label{eq:moy-rels-sym-e}
& \begin{picture}(300,60)(-30,-10)
    \linethickness{0.4mm} \thicklines
    \put(0,0) {
      \linethickness{0.4mm} \thicklines
      \qbezier(0,0)(0,20)(0,40)
      \put(0,40){\vector(0,1){0}}
      %% \put(40,0){\qbezier(0,0)(-20,20)(0,40)}
      %% \put(40,0){\vector(1,-1){0}}
      \put(-10,10){\linethickness{0.15mm}
        \put(10,20){\line(1,0){20}}
      }
      \put(-10,-10){\linethickness{0.15mm}
        \put(10,20){\line(1,0){20}}
      }
    }
    \put(20,0) {
      \linethickness{0.4mm} \thicklines
      \qbezier(0,0)(0,20)(0,40)
      \put(0,40){\vector(0,1){0}}
      %% \put(40,0){\qbezier(0,0)(-20,20)(0,40)}
      %% \put(40,0){\vector(1,-1){0}}
      \put(-10,0){\linethickness{0.15mm}
        \put(10,20){\line(1,0){20}}
      }
    }
    \put(40,0) {
      \linethickness{0.4mm} \thicklines
      \qbezier(0,0)(0,20)(0,40)
      \put(0,40){\vector(0,1){0}}
    }
    \put(50,17){$-$}
    \put(65,0) {
      \linethickness{0.4mm} \thicklines
      \put(0,0) {
        \linethickness{0.4mm} \thicklines
        \qbezier(0,0)(0,20)(0,40)
        \put(0,40){\vector(0,1){0}}
        %% \put(40,0){\qbezier(0,0)(-20,20)(0,40)}
        %% \put(40,0){\vector(1,-1){0}}
        \put(-10,0){\linethickness{0.15mm}
          \put(10,20){\line(1,0){20}}
        }
      }
      \put(20,0) {
        \linethickness{0.4mm} \thicklines
        \qbezier(0,0)(0,20)(0,40)
        \put(0,40){\vector(0,1){0}}
      }
      \put(40,0) {
        \linethickness{0.4mm} \thicklines
        \qbezier(0,0)(0,20)(0,40)
        \put(0,40){\vector(0,1){0}}
      }
    }
    \put(120,17){$\simeq$}
    \put(140,0) {
      \linethickness{0.4mm} \thicklines
      \put(0,0) {
        \linethickness{0.4mm} \thicklines
        \qbezier(0,0)(0,20)(0,40)
        \put(0,40){\vector(0,1){0}}
        %% \put(40,0){\qbezier(0,0)(-20,20)(0,40)}
        %% \put(40,0){\vector(1,-1){0}}
        \put(-10,0){\linethickness{0.15mm}
          \put(10,20){\line(1,0){20}}
        }
      }
      \put(20,0) {
        \linethickness{0.4mm} \thicklines
        \qbezier(0,0)(0,20)(0,40)
        \put(0,40){\vector(0,1){0}}
        \put(-10,10){\linethickness{0.15mm}
          \put(10,20){\line(1,0){20}}
        }
        \put(-10,-10){\linethickness{0.15mm}
          \put(10,20){\line(1,0){20}}
        }
      }
      \put(40,0) {
        \linethickness{0.4mm} \thicklines
        \qbezier(0,0)(0,20)(0,40)
        \put(0,40){\vector(0,1){0}}
      }
      \put(50,17){$-$}
      \put(65,0) {
        \linethickness{0.4mm} \thicklines
        \put(0,0) {
          \linethickness{0.4mm} \thicklines
          \qbezier(0,0)(0,20)(0,40)
          \put(0,40){\vector(0,1){0}}
        }
        \put(20,0) {
          \linethickness{0.4mm} \thicklines
          \qbezier(0,0)(0,20)(0,40)
          \put(0,40){\vector(0,1){0}}
          \put(-10,0){\linethickness{0.15mm}
            \put(10,20){\line(1,0){20}}
          }
        }
        \put(40,0) {
          \linethickness{0.4mm} \thicklines
          \qbezier(0,0)(0,20)(0,40)
          \put(0,40){\vector(0,1){0}}
        }
      }
    }
  \end{picture}
\ee

Again, using only these relations we are not able to uniquely
determine numeric value of every dimension $D_\gamma$.
In particular, we do not know, what is the value of
  $\begin{picture}(20,10)(-5,-2.5)
    \put(0,0){\circle*{5}}
    \put(5,0){\circle{10}}
\end{picture}$.
But what is the \textit{natural} value of it?

At $q = 1$ we can use matrix model and see, that the value is N (N+1).
So, the most natural guess would be that the quantum value is [N][N+1].

But if then, based on this value, we would try to calculate
the value of the old loop {\it dessin} in antisymmetric picture
(using $[1] \otimes [1] = [1,1] + [2]$) we would get

\begin{equation}
  \begin{picture}(20,10)(-5,-2.5)
    \put(0,0){\circle*{5}}
    \put(5,0){\circle{10}}
  \end{picture}_{\text{antisymm}}
   = [2][N] - \begin{picture}(20,10)(-5,-2.5)
    \put(0,0){\circle*{5}}
    \put(5,0){\circle{10}}
   \end{picture}_{\text{symm}}
   = [2][N] - [N][N+1]
\end{equation}

Classically, at $q = 1$ this value is still equal $- N(N-1)$,
but the quantization is different from $-[N][N-1]$,
which seems most natural from ``antisymmetric'' point of view.

Vice versa, if we take the value $-[N][N-1]$ for antisymmetric
loop diagram and calculate symmetric one based on it, we would
get $[2][N] + [N][N-1]$ -- different quantization, than
the one that seems natural from ``symmetric'' point of view.

Since symmetric and antisymmetric points of view are so
interchangeable (related just by transposition of the Young diagram)
there is \textit{a priori} no reason to prefer one over the other.
But choosing one of them would mean that flip-rule
(\ref{eq:flip-and-1-n-rels}) is not true, whereas it holds for the other.
This is an indication that
\textit{the flip rule can not be a consequence of topological invariance}.
It is some extra structure, that is related to preferring one
channel of representation theory over another, in other words,
to symmetry breaking.

This is where we would like to stop this very intriguing
inquiry into representation theory, for now.

\end{document}